\documentstyle[12pt]{article}
\def\ie{{\em i.e.}}
\def\eg{{\em e.g.}}

\def\Zbb{Z\rightarrow b\ov b}
\def\bsg{B(b\to s\gamma)}
\def\epsb{\epsilon_b}

\def\beq{\begin{equation}}
\def\eeq{\end{equation}}

\catcode`\@=11 % This allows us to modify PLAIN macros.
\def\coeff#1#2{{\textstyle{#1\over #2}}}

\def\vev#1{\left\langle #1\right\rangle}
\def\lsim{\mathrel{\mathpalette\@versim<}}
\def\gsim{\mathrel{\mathpalette\@versim>}}
\def\@versim#1#2{\vcenter{\offinterlineskip
    \ialign{$\m@th#1\hfil##\hfil$\crcr#2\crcr\sim\crcr } }}
\def\etal{{\em et. al.}}
\def\JL{J. L. Lopez}
\def\DVN{D. V. Nanopoulos}
\def\AZ{A. Zichichi}

\def\XW{X. Wang}

\def\t1{{\tilde 1}}
\def\ov{\overline}

\def\GeV{\,{\rm GeV}}
\def\TeV{\,{\rm TeV}}

\def\wt{\widetilde}

\def\to{\rightarrow}
\def\pb{\,{\rm pb}}
\def\fb{\,{\rm fb}}
\def\ipb{\,{\rm pb}^{-1}}
\def\NPB#1#2#3{Nucl. Phys. B {\bf#1} (19#2) #3}
\def\PLB#1#2#3{Phys. Lett. B {\bf#1} (19#2) #3}

\def\PRD#1#2#3{Phys. Rev. D {\bf#1} (19#2) #3}
\def\PRL#1#2#3{Phys. Rev. Lett. {\bf#1} (19#2) #3}
\def\PRT#1#2#3{Phys. Rep. {\bf#1} (19#2) #3}
\def\MODA#1#2#3{Mod. Phys. Lett. A {\bf#1} (19#2) #3}

\def\TAMU#1{Texas A \& M University preprint CTP-TAMU-#1}

\begin{document}

% TH format
\begin{flushright}
\baselineskip=12pt
{CERN-TH.7139/94}\\
{CERN-LAA/94-01}\\
{CTP-TAMU-74/93}\\
{ACT-26/93}\\
\end{flushright}
% PPE format
%\begin{center}
%{\large EUROPEAN ORGANIZATION FOR NUCLEAR RESEARCH}
%\end{center}
%\begin{flushright}
%{CERN-PPE/93-??}\\
%{?? July, 1993}\\
%{CERN-LAA/93-??}\\
%{CERN-TH.????/93}\\
%{CTP-TAMU-40/93}\\
%{ACT-15/93}\\
%\end{flushright}
%

\begin{center}
%\vglue 0.3cm
{\Large\bf Experimental Aspects of SU(5)xU(1) Supergravity\\}
\vspace{0.2cm}
\vglue 0.5cm
{JORGE L. LOPEZ$^{(a),(b)}$, D. V. NANOPOULOS$^{(a),(b),(c)}$,
GYE T. PARK$^{(a),(b)}$,\\}
{XU WANG$^{(a),(b)}$, and A. ZICHICHI$^{(d)}$\\}
\vglue 0.2cm
{\em $^{(a)}$Center for Theoretical Physics, Department of Physics, Texas A\&M
University\\}
{\em College Station, TX 77843--4242, USA\\}
{\em $^{(b)}$Astroparticle Physics Group, Houston Advanced Research Center
(HARC)\\}
{\em The Mitchell Campus, The Woodlands, TX 77381, USA\\}
{\em $^{(c)}$CERN, Theory Division, 1211 Geneva 23, Switzerland\\}
{\em $^{(d)}$CERN, 1211 Geneva 23, Switzerland\\}
\baselineskip=12pt

\vglue 0.5cm
{\tenrm ABSTRACT}
\end{center}
\vglue -0.2cm
{\rightskip=3pc
 \leftskip=3pc
%\xpt\baselineskip=12pt
\noindent
We study various aspects of $SU(5)\times U(1)$ supergravity as they relate
to the experimental verification or falsification of this model. We consider
two string-inspired, universal, one-parameter, no-scale
soft-supersymmetry-breaking scenarios, driven by the $F$-terms of the moduli
and dilaton fields. The model is described in terms of the supersymmetry mass
scale (\ie, the chargino mass $m_{\chi^\pm_1}$), $\tan\beta$, and the top-quark
mass. We first determine the combined effect on the parameter space of all
presently available direct and indirect experimental constraints, including the
LEP lower bounds on sparticle and Higgs-boson masses, the $b\to s\gamma$ rate,
the anomalous magnetic moment of the muon, the high-precision electroweak
parameters $\epsilon_1,\epsilon_b$ (which imply $m_t\lsim180\GeV$), and the
muon fluxes in underground detectors (neutrino telescopes). For the
still-allowed points in $(m_{\chi^\pm_1},\tan\beta)$ parameter space, we
re-evaluate the experimental situation at the Tevatron, LEPII, and HERA. In the
1994 run, the Tevatron could probe chargino masses as high as 100 GeV. At LEPII
the parameter space could be explored with probes of different resolutions:
Higgs boson searches, selectron searches, and chargino searches. Moreover, for
$m_t\lsim150\GeV$, these Higgs-boson searches could explore all of the allowed
parameter space with $\sqrt{s}\lsim210\GeV$.}

% TH format
\vspace{0.5cm}
\begin{flushleft}
\baselineskip=12pt
{CERN-TH.7139/94}\\
{CERN-LAA/94-01}\\
{CTP-TAMU-74/93}\\
{ACT-26/93}\\
December 1993
\end{flushleft}
\vfill\eject
\setcounter{page}{1}
\pagestyle{plain}

\baselineskip=14pt

\section{Introduction}

In the search for physics beyond the Standard Model, what is needed
are detailed calculations to be confronted with experimental data. The
starting point is the choice of a model described by the least numbers of
parameters, and based on well motivated theoretical assumptions. Our choice
is $SU(5)\times U(1)$ supergravity \cite{EriceDec92}, the reasons being
two-fold. First, because this model is derivable from string theory. Second,
because the $SU(5)\times U(1)$ gauge group is the simplest unified gauge
extension of the Standard Model. It is unified because the two non-abelian
gauge couplings of the Standard Model ($\alpha_2$ and $\alpha_3$)
are unified into the $SU(5)$ gauge coupling. It is the simplest
extension because this is the smallest unified group which provides
neutrino masses. In this interpretation, minimal $SU(5)$ would
appear as a subgroup of $SO(10)$, if it is to allow for neutrino masses.
Moreover, the matter representations of $SU(5)\times U(1)$ entail
several simplifications \cite{revitalized}. The most important are: (i) the
breaking of the gauge group via vacuum expectation values of
$10,\overline{10}$ Higgs fields; (ii) the natural splitting of the
doublet and triplet components of the Higgs pentaplets and therefore the
natural avoidance of dangerous dimension-five proton decay operators; and
(iii) the natural appearance of a see-saw mechanism for neutrino
masses. In the context of string model-building, the $SU(5)\times U(1)$
structure becomes even more important, since the traditional grand
unified gauge groups ($SU(5), SO(10), E_6$) cannot be broken down to the
Standard Model gauge group in the simplest (and to date almost unique)
string constructions, because of the absence of adjoint Higgs
representations \cite{ELN}. This reasoning is not applicable to the
$SU(5)\times U(1)$ gauge group, since the required $10,\overline{10}$
representations are very common in string model-building
\cite{revamped,decisive,LNY}.

We supplement the $SU(5)\times U(1)$ gauge group choice with the minimal
matter content which allows it to unify at the string scale $M_U \sim 10^{18}
\GeV$, as expected to occur in the string-derived versions of the model
\cite{Lacaze,thresholds}. This entails a set of intermediate-scale mass
particles: a vector-like quark doublet with mass $m_Q \sim 10^{12} \GeV$ and a
vector-like charge $-1/3$ quark singlet with mass $m_D \sim 10^6 \GeV$
\cite{sism,LNZI}. The model is also implicitly constrained by the requirement
of suitable supersymmetry breaking. We choose two string-inspired scenarios
which have the virtue of yielding universal soft-supersymmetry-breaking
parameters ($m_{1/2},m_0,A$), in contrast with non-universal
soft-supersymmetry-breaking scenarios which occur quite commonly in string
constructions \cite{IL,KL,Ibanez} and may be phenomenologically troublesome
\cite{EN}. These scenarios are example of the no-scale supergravity framework
\cite{Lahanas+EKNI+II,LN} in which the dimensional parameters of the theory are
undetermined at the classical level, but are fixed by radiative corrections,
thus including the whole theory in the determination of the low-energy
parameters. In the {\em moduli} scenario, supersymmetry breaking is driven by
the vev of the moduli fields ($T$), and gives $m_0 = A = 0$, while in the {\em
dilaton} scenario \cite{KL,Ibanez} supersymmetry breaking is driven by the vev
of the dilaton field ($S$) and entails $m_0 = \frac{1}{\sqrt{3}}m_{1/2},\
A=-m_{1/2}$. Thus, the supersymmetry breaking sector depends on only one
parameter ({\em i.e.}, $m_{1/2}$).

The parameter space of $SU(5)\times U(1)$ supergravity is fully described
by just two more quantities: the ratio of Higgs-boson vacuum expectation values
($\tan\beta$), and the top-quark mass ($m_t$). This three-dimensional parameter
space ({\em i.e.}, $m_{1/2}, \tan\beta$ and $m_t$) has been explored in detail
in Refs.~\cite{LNZI} and \cite{LNZII} for the moduli and dilaton scenarios
respectively. The allowed points in parameter space are determined by a
theoretical procedure (including renormalization group evolution of the model
parameters from the unification scale down to the electroweak scale and
enforcement of radiative electroweak symmetry breaking using the one-loop
effective potential) and by the further imposition of the basic LEP
constraints on the sparticle and Higgs-boson masses, as described in
Ref.~\cite{aspects}.  More recently, we have investigated further constraints
on the parameter space, including: (i) the CLEO limits on the $b\to s\gamma$
rate \cite{bsgamma,bsg-eps}, (ii) the long-standing limit on the anomalous
magnetic moment of the muon \cite{g-2}, (iii) the electroweak high-precision
LEP measurements in the form of the $\epsilon_{1},\epsilon_b$ parameters
\cite{ewcorr,bsg-eps,eps1-epsb} (here we update our analysis including the
latest LEP data), (iv) the non-observation of anomalous muon fluxes in
underground detectors (``neutrino telescopes") \cite{NT}, and (v) the possible
constraints from trilepton searches at the Tevatron \cite{LNWZ}.

In our analysis we combine the most useful elements of the top-down and
bottom-up approaches to physics beyond the Standard Model. The top-down
approach consists of selecting particularly well motivated string-inspired
scenarios for supersymmetry breaking ({\em i.e.}, with a single mass
parameter), whereas the bottom-up approach aims at imposing all known direct
and indirect experimental constraints on the chosen model. In this way, we can
corner the high-energy parameter space of the model (bottom-up) and thus focus
our search for further realistic supersymmetric models (top-down). On the other
hand, the completely phenomenological approach in which the many parameters
(more than 20) of the Minimal Supersymmetric Standard Model (MSSM) are
arbitrarily varied, is neither practical nor illuminating.

It is important to note that our advocacy of supersymmetry, as the choice for
physics beyond the Standard Model, seems to be accumulating indirect supporting
evidence: (i) global fits to the electroweak sector of the Standard Model show
a preference for a light Higgs boson \cite{EF}, in agreement with low-energy
supersymmetry where a light Higgs boson is always present; (ii) the precisely
measured gauge couplings, when extrapolated to very high energies using
Standard Model radiative effects, fail to converge at any high-energy scale
\cite{oldAmaldi,EKN}, consistent with the fate of non-supersymmetric GUTs in
light of the gauge hierarchy problem; (iii) on the contrary, in the
supersymmetric version of the Standard Model, the gauge couplings unify at a
scale $M_U \sim 10^{16} \GeV$ \cite{EKN}; (iv) global fits to the electroweak
data also imply that $m_t = 140\pm20 \GeV$ for $m_H = 60 \GeV$ and $m_t =
180\pm18 \GeV$ for $m_H = 1 \TeV$ (see {\em e.g.}, Ref.~\cite{Altarelli,EFL}),
consistent with the radiative electroweak symmetry breaking mechanism
\cite{EWx,LN}; (v) $m_t \lsim 190-200 \GeV$ (see {\em e.g.}, Ref.~\cite{DL}) is
required in a supersymmetric unified theory, consistent with the electroweak
fits to $m_t$; and (vi) the resulting top-quark Yukawa couplings at the
unification scale are naturally obtained in supersymmetric string models
\cite{revamped,decisive}.

In this paper we first briefly review the basic $SU(5)\times U(1)$ supergravity
properties (section~\ref{model}), and then discuss each of the constraints on
the parameter space separately (sections~\ref{epsilons},\ref{constraints}), and
also their combined effect (section~\ref{allowed}). Next we address the
prospects for detecting the sparticles and Higgs bosons directly through
searches at the Tevatron, LEPII, and HERA (section~\ref{direct}). We conclude
that with the present generation of collider facilities, direct searches for
the lighter weakly interacting sparticles and Higgs bosons probe the parameter
space of $SU(5)\times U(1)$ supergravity in a much deeper way than direct
searches for the heavier strongly interacting sparticles do. Moreover, within
the weakly interacting sparticles, the deepest probe is provided by the
lightest Higgs boson, followed by the selectrons, and then by the charginos. We
also discuss the two most efficient ways of exploring the parameter space in
the near future in an indirect way (section~\ref{indirect}), namely through
more precise $B(b\to s\gamma)$ and $(g-2)_\mu$ measurements. We summarize our
conclusions in section~\ref{conclusions}.

\section{SU(5)xU(1) supergravity}
\label{model}
\subsection{Model building}
The supergravity model of interest is based on the gauge group $SU(5)\times
U(1)$ and is best motivated as a possible solution to string theory. In this
regard several of its features become singularly unique, as discussed in the
Introduction. However, honest-to-goodness string models (such as the one in
Ref.~\cite{LNY}) are quite complicated and their phenomenology tends to be
obscured by a number of new string parameters (although these could in
principle be determined dynamically). It is therefore more convenient to study
the phenomenology of a ``string-inspired" model \cite{LNZI} which contains all
the desirable features of the real string model, but where several simplifying
assumptions have been made, as ``inspired" by the detailed calculations in the
real model. The string-inspired model is such that unification of the
low-energy gauge couplings of the Standard Model occurs at the string scale
$M_U\sim10^{18}\GeV$. This is a simplifying assumption since in the string
model there are several intermediate-scale particles which in effect
produce a threshold structure as the string scale is approached. Perhaps
because of this simplifying assumption, in the string-inspired model one seems
to be forced to introduce non-minimal matter representations at intermediate
scales: a vector-like quark doublet with mass $m_Q\sim10^{12}\GeV$ and
a vector-like charge $-1/3$ quark singlet with mass $m_D\sim10^6\GeV$
\cite{sism,LNZI}. The low-energy spectrum of the model contains the same
sparticles and Higgs bosons as the Minimal Supersymmetric Standard Model
(MSSM).

A very important component of the model is that which triggers supersymmetry
breaking. In the string model this task is performed by the hidden
sector and the universal moduli and dilaton fields. Model-dependent
calculations are required to determine the precise nature of supersymmetry
breaking in a given string model. In fact, no explicit string model exists to
date where various theoretical difficulties (\eg, suitably suppressed
cosmological constant, suitable vacuum state with perturbative gauge coupling,
etc.) have been satisfactorily overcome. Instead, it has become
apparent \cite{IL,KL,Ibanez} that a more model-independent approach to the
problem may be more profitable. In this approach one parametrizes the breaking
of supersymmetry by the largest $F$-term vacuum expectation value which
triggers supersymmetry breaking. Of all the possible fields which could be
involved (\ie, hidden sector matter fields, various moduli fields, dilaton)
the dilaton and three of the moduli fields are quite common in string
constructions and have thus received the most attention in the literature. In a
way, if supersymmetry breaking is triggered by these fields (\ie,
$\vev{F_S}\not=0$ or $\vev{F_T}\not=0$), this would be a rather generic
prediction of string theory.

There are various possible scenarios for supersymmetry breaking that are
obtained in this model-independent way. To discriminate among these we
consider a simplified expression for the scalar masses (\eg, $m_{\tilde q}$)
$\widetilde m^2_i=m^2_{3/2}(1+n_i\cos^2\theta)$, with
$\tan\theta=\vev{F_S}/\vev{F_T}$ \cite{Ibanez}. Here $m_{3/2}$ is the gravitino
mass and the $n_i$ are the modular weights of the respective matter field.
There are two ways in which one can obtain universal scalar masses, as desired
phenomenologically to avoid large flavor-changing-neutral-currents (FCNCs)
\cite{EN}: (i) setting $\theta=\pi/2$, that is $\vev{F_S}\gg\vev{F_T}$; or (ii)
in a model where all $n_i$ are the same, as occurs for $Z_2\times Z_2$
orbifolds \cite{Ibanez} and free-fermionic constructions \cite{thresholds}.

In the first (``dilaton") scenario, supersymmetry breaking is triggered by the
dilaton $F$-term and yields universal soft-supersymmetry-breaking gaugino and
scalar masses and trilinear interactions \cite{KL,Ibanez}
\beq
m_0=\coeff{1}{\sqrt{3}}m_{1/2},\qquad A=-m_{1/2}.\label{dilaton}
\eeq
In the second (``moduli") scenario, in the limit $\vev{F_T}\gg\vev{F_S}$ (\ie,
$\theta\to0$) all scalar masses at the unification scale vanish, as is the case
in no-scale supergravity models with a unified group structure \cite{LN}. In
this case we have
\beq
m_0=0,\qquad A=0. \label{noscale}
\eeq

The procedure to extract the low-energy predictions of the model outlined
above is rather standard (see \eg, Ref. \cite{aspects}): (a) the bottom-quark
and tau-lepton masses, together with the input values of $m_t$ and $\tan\beta$
are used to determine the respective Yukawa couplings at the electroweak scale;
(b) the gauge and Yukawa couplings are then run up to the unification scale
$M_U=10^{18}\GeV$ taking into account the intermediate-scale particles
introduced above; (c) at the unification scale the
soft-supersymmetry-breaking parameters are introduced (according to Eqs.
(\ref{dilaton},\ref{noscale})) and the scalar masses are then run down to the
electroweak scale; (d) radiative electroweak symmetry breaking is enforced by
minimizing the one-loop effective potential which depends on the whole mass
spectrum, and the values of the Higgs mixing term $|\mu|$ and the bilinear
soft-supersymmetry breaking parameter $B$ are determined from the minimization
conditions; (e) all known phenomenological constraints on the sparticle and
Higgs-boson masses are applied (most importantly the LEP lower bounds on the
chargino and Higgs masses), including the cosmological requirement of a
not-too-large neutralino relic density (which happens to be satisfied
automatically).

In either of the supersymmetry breaking scenarios considered, after enforcement
of the above constraints, the low-energy theory can be described in terms of
just three parameters: the top-quark mass ($m_t$), the ratio of Higgs vacuum
expectation values ($\tan\beta$), and the gaugino mass ($m_{1/2}$). Therefore,
measurement of only two sparticle or Higgs-boson masses would determine the
remaining thirty. Moreover, if the hidden sector responsible for these patterns
of soft supersymmetry breaking is specified (as in a string-derived model),
then the gravitino mass will also be determined and the supersymmetry breaking
sector of the theory will be completely fixed.

\subsection{Mass ranges}
We have scanned the three-dimensional parameter space for
$m_t=130,150,170\GeV$, $\tan\beta=2\to50$ and $m_{1/2}=50\to500\GeV$. Imposing
the constraint $m_{\tilde g},m_{\tilde q}<1\TeV$ we find
\begin{eqnarray}
&{\rm moduli}:\qquad	&m_{1/2}<475\GeV,\quad \tan\beta\lsim32,\\
&{\rm dilaton}:\qquad	&m_{1/2}<465\GeV,\quad \tan\beta\lsim46.
\end{eqnarray}
These restrictions on $m_{1/2}$ cut off the growth of most of the sparticle and
Higgs masses at $\approx1\TeV$. However, the sleptons, the lightest Higgs
boson, the two lightest neutralinos, and the lightest chargino are cut off at a
much lower mass, as follows\footnote{In this class of supergravity models the
three sneutrinos ($\tilde\nu$) are degenerate in mass. Also, $m_{\tilde
\mu_L}=m_{\tilde e_L}$ and $m_{\tilde\mu_R}=m_{\tilde e_R}$.}
\begin{eqnarray}
&{\rm moduli}:&\left\{
	\begin{array}{l}
	m_{\tilde e_R}<190\GeV,\quad m_{\tilde e_L}<305\GeV,
				\quad m_{\tilde\nu}<295\GeV\\
	m_{\tilde\tau_1}<185\GeV,\quad m_{\tilde\tau_2}<315\GeV\\
	m_h<125\GeV\\
	m_{\chi^0_1}<145\GeV,\quad m_{\chi^0_2}<290\GeV,
				\quad m_{\chi^\pm_1}<290\GeV
	\end{array}
		\right.\\
&{\rm dilaton}:&\left\{
	\begin{array}{l}
	m_{\tilde e_R}<325\GeV,\quad m_{\tilde e_L}<400\GeV,
				\quad m_{\tilde\nu}<400\GeV\\
	m_{\tilde\tau_1}<325\GeV,\quad m_{\tilde\tau_2}<400\GeV\\
	m_h<125\GeV\\
	m_{\chi^0_1}<145\GeV,\quad m_{\chi^0_2}<285\GeV,
				\quad m_{\chi^\pm_1}<285\GeV
	\end{array}
		\right.
\end{eqnarray}
It is interesting to note that because of the various constraints on the model,
the gluino and (average) squark masses are bounded from below,
\beq
{\rm moduli}:\left\{
	\begin{array}{l}
	m_{\tilde g}\gsim245\,(260)\GeV\\
	m_{\tilde q}\gsim240\,(250)\GeV
	\end{array}
		\right.
\qquad
{\rm dilaton}:\left\{
	\begin{array}{l}
	m_{\tilde g}\gsim195\,(235)\GeV\\
	m_{\tilde q}\gsim195\,(235)\GeV
	\end{array}
		\right.		\label{gmin}
\eeq
for $\mu>0(\mu<0)$. Relaxing the above conditions on $m_{1/2}$ simply allows
all sparticle masses to grow further proportional to $m_{\tilde g}$.

\subsection{Mass relations}
The neutralino and chargino masses show a correlation observed before in
this class of models \cite{ANc,LNZI}, namely (see Fig.~\ref{Figure1}, top row)
\beq
m_{\chi^0_1}\approx \coeff{1}{2}m_{\chi^0_2},\qquad
m_{\chi^0_2}\approx m_{\chi^\pm_1}\approx M_2=(\alpha_2/\alpha_3)m_{\tilde g}
\approx0.28m_{\tilde g}.\label{neuchar}
\eeq
This is because throughout the parameter space $|\mu|$ is generally much larger
than $M_W$ (see Fig.~\ref{Figure1}, bottom row) and $|\mu|>M_2$. In practice we
find $m_{\chi^0_2}\approx m_{\chi^\pm_1}$ to be satisfied quite accurately,
whereas $m_{\chi^0_1}\approx{1\over2}m_{\chi^0_2}$ is only qualitatively
satisfied, although the agreement is better in the ${\rm dilaton}$ case. In
fact, these two mass relations are much more reliable than the one that links
them to $m_{\tilde g}$. The heavier neutralino ($\chi^0_{3,4}$) and chargino
($\chi^\pm_2$) masses are determined by the value of $|\mu|$; they all approach
this limit for large enough $|\mu|$. More precisely, $m_{\chi^0_3}$ approaches
$|\mu|$ sooner than $m_{\chi^0_4}$ does. On the other hand, $m_{\chi^0_4}$
approaches $m_{\chi^\pm_2}$ rather quickly.

The first- and second-generation squark and slepton masses can be determined
analytically
\beq
\wt m_i=\left[m^2_{1/2}(c_i+\xi^2_0)-d_i{\tan^2\beta-1\over\tan^2\beta+1}
M^2_W\right]^{1/2}=a_i m_{\tilde g}\left[1+b_i\left({150\over m_{\tilde
g}}\right)^2{\tan^2\beta-1\over\tan^2\beta+1}\right]^{1/2},\label{masses}
\eeq
where $d_i=(T_{3i}-Q)\tan^2\theta_w+T_{3i}$ (\eg, $d_{\tilde u_L}={1\over2}
-{1\over6}\tan^2\theta_w$, $d_{\tilde e_R}=-\tan^2\theta_w$), and
$\xi_0=m_0/m_{1/2}=0,\coeff{1}{\sqrt{3}}$. The coefficients $c_i$ can be
calculated numerically in terms of the low-energy gauge couplings, and are
given in  Table~\ref{Table1}\footnote{These are renormalized at the scale
$M_Z$. In a more accurate treatment, the $c_i$ would be renormalized at the
physical sparticle mass scale, leading to second order shifts on the sparticle
masses.} for $\alpha_3(M_Z)=0.118\pm0.008$. In the table we also give
$c_{\tilde g}=m_{\tilde g}/m_{1/2}$. Note that these values are smaller than
what is obtained in the minimal $SU(5)$ supergravity model (where $c_{\tilde
g}=2.90$ for $\alpha_3(M_Z)=0.118$) and therefore the numerical relations
between the gluino mass and the neutralino masses are different in that model.
In the table we also show the resulting values for $a_i,b_i$ for the central
value of $\alpha_3(M_Z)$.

\begin{table}
\hrule
\caption{
The value of the $c_i$ coefficients appearing in  Eq.~(9), the ratio $c_{\tilde
g}=m_{\tilde g}/m_{1/2}$, and the average squark coefficient
$\bar c_{\tilde q}$, for $\alpha_3(M_Z)=0.118\pm0.008$. Also shown are the
$a_i,b_i$ coefficients for the central value of $\alpha_3(M_Z)$ and both
supersymmetry breaking scenaria ($T$: moduli, $S$: dilaton). The results apply
as well to the second-generation squark and slepton masses.}
\label{Table1}
\begin{center}
\begin{tabular}{|c|c|c|c|}\hline
$i$&$c_i\,(0.110)$&$c_i\,(0.118)$&$c_i\,(0.126)$\\ \hline
$\tilde\nu,\tilde e_L$&$0.406$&$0.409$&$0.413$\\
$\tilde e_R$&$0.153$&$0.153$&$0.153$\\
$\tilde u_L,\tilde d_L$&$3.98$&$4.41$&$4.97$\\
$\tilde u_R$&$3.68$&$4.11$&$4.66$\\
$\tilde d_R$&$3.63$&$4.06$&$4.61$\\
$c_{\tilde g}$&$1.95$&$2.12$&$2.30$\\
$\bar c_{\tilde q}$&$3.82$&$4.07$&$4.80$\\ \hline
\end{tabular}
%\quad
\begin{tabular}{|c|c|c|c|c|}\hline
$i$&$a_i(T)$&$b_i(T)$&$a_i(S)$&$b_i(S)$\\ \hline
$\tilde e_L$&$0.302$&$+1.115$&$0.406$&$+0.616$\\
$\tilde e_R$&$0.185$&$+2.602$&$0.329$&$+0.818$\\
$\tilde\nu$&$0.302$&$-2.089$&$0.406$&$-1.153$\\
$\tilde u_L$&$0.991$&$-0.118$&$1.027$&$-0.110$\\
$\tilde u_R$&$0.956$&$-0.016$&$0.994$&$-0.015$\\
$\tilde d_L$&$0.991$&$+0.164$&$1.027$&$+0.152$\\
$\tilde d_R$&$0.950$&$-0.033$&$0.989$&$-0.030$\\ \hline
\end{tabular}
\end{center}
\hrule
\end{table}

The ``average" squark mass, $m_{\tilde q}\equiv{1\over8}(m_{\tilde
u_L}+m_{\tilde u_R}+m_{\tilde d_L}+m_{\tilde d_R}+m_{\tilde c_L}+m_{\tilde c_R}
+m_{\tilde s_L}+m_{\tilde s_R})
=(m_{\tilde g}/c_{\tilde q})\sqrt{\bar c_{\tilde q}+\xi^2_0}$, with $\bar
c_{\tilde q}$ given in Table~\ref{Table1}, is determined to be
\beq
m_{\tilde q}=\left\{	\begin{array}{ll}
			(1.00,0.95,0.95) m_{\tilde g},&\quad{\rm moduli}\\
			(1.05,0.99,0.98) m_{\tilde g},&\quad{\rm dilaton}
			\end{array}
		\right.
\eeq
for $\alpha_3(M_Z)=0.110,0.118,0.126$ (the dependence on $\tan\beta$ is small).
The squark splitting around the average is $\approx2\%$.

The first- and second-generation squark and slepton masses are plotted in
Fig.~\ref{Figure2}. The thickness and straightness of the lines shows the small
$\tan\beta$ dependence, except for $\tilde\nu$. The results do not depend on
the sign of $\mu$, except to the extent that some points in parameter space are
not allowed for both signs of $\mu$: the $\mu<0$ lines start-off at larger mass
values. Note that
\beq
{\rm moduli}:\left\{
	\begin{array}{l}
	m_{\tilde e_R}\approx0.18m_{\tilde g}\\
	m_{\tilde e_L}\approx0.30m_{\tilde g}\\
	m_{\tilde e_R}/m_{\tilde e_L}\approx0.61
	\end{array}
		\right.
\qquad
{\rm dilaton}:\left\{
	\begin{array}{l}
	m_{\tilde e_R}\approx0.33m_{\tilde g}\\
	m_{\tilde e_L}\approx0.41m_{\tilde g}\\
	m_{\tilde e_R}/m_{\tilde e_L}\approx0.81
	\end{array}
		\right.
\eeq

The third generation squark and slepton masses cannot be determined
analytically. These are shown in Fig.~\ref{Figure3}, and exhibit a large
variability for fixed $m_{\tilde g}$ because of the $\tan\beta$-dependence in
the off-diagonal element of the corresponding $2\times2$ mass matrices. The
lowest values of the $\tilde t_1$ mass go up with $m_t$ and can be as low as
\beq
m_{\tilde t_1}\gsim\left\{	\begin{array}{ll}
		160,170,190\,(155,150,170)\GeV;&\quad{\rm moduli}\\
			88,112,150\,(92,106,150)\GeV;&\quad{\rm dilaton}
			\end{array}
		\right.
\eeq
for $m_t=130,150,170\GeV$ and $\mu>0\,(\mu<0)$.

The one-loop corrected lightest CP-even ($h$) and CP-odd ($A$) Higgs boson
masses are shown in Fig.~\ref{Figure4}. Following the methods of Ref.
\cite{LNPWZh} we have determined that the LEP lower bound on $m_h$ becomes
$m_h\gsim60\GeV$. The largest value of $m_h$ depends on $m_t$; we find
\beq
m_h<\left\{	\begin{array}{ll}
		106,115,125\GeV;&\quad{\rm moduli}\\
		107,117,125\GeV;&\quad{\rm dilaton}
			\end{array}
		\right.
\eeq
for $m_t=130,150,170\GeV$. Note that even though $m_A$ can be fairly light, we
always get $m_A>m_h$, in agreement with a general theorem to this effect in
supergravity theories \cite{DNh}. This result also implies that the channel
$e^+e^-\to hA$ at LEPI is not kinematically allowed in this model.

The computation of the neutralino relic density (following the methods of
Refs. \cite{LNYdmI,KLNPYdm}) shows that $\Omega_\chi h^2_0\lsim0.25\,(0.90)$ in
the moduli (dilaton) scenarios. This implies that in these models the
cosmologically interesting values $\Omega_\chi h^2_0\lsim1$ occur quite
naturally. These results are in good agreement with the observational upper
bound on $\Omega_\chi h^2_0$ \cite{KT}.

\begin{table}
\hrule
\caption{The approximate proportionality coefficients to the gluino mass, for
the various sparticle masses in the two supersymmetry breaking scenarios
considered. The $|\mu|$ coefficients apply for $m_t=150\GeV$ only.}
\label{Table2}
\begin{center}
\begin{tabular}{|c|c|c|}\hline
&moduli&dilaton\\ \hline
$\tilde e_R,\tilde \mu_R$&$0.18$&$0.33$\\
$\tilde\nu$&$0.18-0.30$&$0.33-0.41$\\
$2\chi^0_1,\chi^0_2,\chi^\pm_1$&$0.28$&$0.28$\\
$\tilde e_L,\tilde \mu_L$&$0.30$&$0.41$\\
$\tilde q$&$0.97$&$1.01$\\
$\tilde g$&$1.00$&$1.00$\\
$|\mu|$&$0.5-0.7$&$0.6-0.8$\\\hline
\end{tabular}
\end{center}
\hrule
\end{table}

As we have discussed, in the scenarios we consider all sparticle masses scale
with the gluino mass, with a mild $\tan\beta$ dependence (except for the
third-generation squark and slepton masses). In Table~\ref{Table2} we collect
the approximate proportionality coefficients to the gluino mass for each
sparticle mass (not including the third generation squarks and sleptons).
{}From this table one can (approximately) translate any bounds on a given
sparticle mass on bounds on all the other sparticle masses.

\subsection{Special cases}
\subsubsection{The strict no-scale case}
We now impose the additional constraint $B(M_U)=0$ to be added to
Eq.~(\ref{noscale}), and obtain the so-called strict no-scale case \cite{LNZI}.
Since $B(M_Z)$ is determined by the radiative electroweak symmetry breaking
conditions, this added constraint needs to be imposed in a rather indirect way.
That is, for given $m_{\tilde g}$ and $m_t$ values, we scan the possible values
of $\tan\beta$ looking for cases where $B(M_U)=0$. The most striking result is
that solutions exist {\em only} for $m_t\lsim135\GeV$ if $\mu>0$ and for
$m_t\gsim140\GeV$ if $\mu<0$. That is, the value of $m_t$ {\em determines} the
sign of $\mu$. Furthermore, for $\mu<0$ the value of $\tan\beta$ is determined
uniquely as a function of $m_t$ and $m_{\tilde g}$, whereas for $\mu>0$,
$\tan\beta$ can be double-valued for some $m_t$ range which includes
$m_t=130\GeV$.

All the mass relationships deduced in the previous subsection apply here as
well. The $\tan\beta$-spread that some of them have will be much reduced
though. The most noticeable changes occur for the quantities which depend most
sensitively on $\tan\beta$, such as the Higgs-boson masses. Figure 5 of
Ref.~\cite{EriceDec92} shows that the one-loop corrected lightest Higgs-boson
mass is largely determined by $m_t$, with a weak dependence on $m_{\tilde g}$.
Moreover, for $m_t\lsim135\GeV\Leftrightarrow\mu>0$, $m_h\lsim105\GeV$; whereas
for $m_t\gsim140\GeV\Leftrightarrow\mu<0$, $m_h\gsim100\GeV$. Therefore, in the
strict no-scale case, once the top-quark mass is measured, we will know the
sign of $\mu$ and whether $m_h$ is above or below $100\GeV$.

\subsubsection{The special dilaton scenario case}
\label{specialdilaton}
In the analysis described above, the radiative electroweak breaking conditions
were used to determine the magnitude of the Higgs mixing term $\mu$ at the
electroweak scale. This quantity is ensured to remain light as long as the
supersymmetry breaking parameters remain light. In a fundamental theory this
parameter should be calculable and its value used to determine the $Z$-boson
mass. From this point of view it is not clear that the natural value of $\mu$
should be light. In specific models one can obtain such values by invoking
non-renormalizable interactions \cite{muproblem,Casasmu,decisive}. Another
contribution to this quantity is generically present in string supergravity
models \cite{GM,Casasmu,KL}. The general case with contributions from both
sources has been effectively dealt with in the previous section. If one assumes
that only supergravity-induced contributions to $\mu$ exist, then it can be
shown that the $B$-parameter at the unification scale is also determined
\cite{KL,Ibanez},
\beq
B(M_U)=2m_0=\coeff{2}{\sqrt{3}}m_{1/2},\label{klII}
\eeq
which is to be added to the set of relations in Eq. (\ref{dilaton}). This new
constraint effectively determines $\tan\beta$ for given $m_t$ and $m_{\tilde
g}$ values and makes this restricted version of the model highly predictive
\cite{LNZII}.

It can be shown \cite{LNZII} that only solutions with $\mu<0$ exist. A
numerical iterative procedure allows us to determine the value of $\tan\beta$
which satisfies Eq.~(\ref{klII}), from the calculated value of $B(M_Z)$. We
find that
\beq
\tan\beta\approx1.57-1.63,1.37-1.45,1.38-1.40\quad{\rm for\ }
m_t=130,150,155\GeV
\eeq
is required. Since $\tan\beta$ is so small ($m^{tree}_h\approx28-41\GeV$), a
significant one-loop correction to $m_h$ is required to increase it above
its experimental lower bound of $\approx60\GeV$ \cite{LNPWZh}. This requires
the largest possible top-quark masses and a not-too-small squark mass. However,
perturbative unification imposes an upper bound on $m_t$ for a given
$\tan\beta$ \cite{DL}, which in this case implies \cite{aspects}
\beq
m_t\lsim155\GeV,
\eeq
which limits the magnitude of $m_h$
\beq
m_h\lsim74,87,91\GeV\qquad{\rm for}\qquad m_t=130,150,155\GeV.
\eeq
In Table~\ref{Table3} we give the range of sparticle and Higgs masses that
are allowed in this case.

\begin{table}
\hrule
\caption{
The range of allowed sparticle and Higgs-boson masses in the special dilaton
scenario. The top-quark mass is restricted to be $m_t<155\GeV$. All masses in
GeV.}
\label{Table3}
\begin{center}
\begin{tabular}{|c|c|c|c|}\hline
$m_t$&$130$&$150$&$155$\\ \hline
$\tilde g$&$335-1000$&$260-1000$&$640-1000$\\
$\chi^0_1$&$38-140$&$24-140$&$90-140$\\
$\chi^0_2,\chi^\pm_1$&$75-270$&$50-270$&$170-270$\\
$\tan\beta$&$1.57-1.63$&$1.37-1.45$&$1.38-1.40$\\
$h$&$61-74$&$64-87$&$84-91$\\
$\tilde l$&$110-400$&$90-400$&$210-400$\\
$\tilde q$&$335-1000$&$260-1000$&$640-1000$\\
$A,H,H^+$&$>400$&$>400$&$>970$\\ \hline
\end{tabular}
\end{center}
\hrule
\end{table}

\section{Updated precision electroweak tests}
\label{epsilons}
Among the various schemes to parametrize the electroweak vacuum polarization
corrections \cite{Kennedy,PT,efflagr,AB}, we choose the so-called
$\epsilon$-scheme \cite{ABJ,ABC} where the model predictions are absolute and
valid to higher orders in $q^2$. This scheme is therefore more applicable to
the electroweak precision tests of the MSSM \cite{BFC} and a class of
supergravity models \cite{ewcorr}. There are two $\epsilon$-schemes. The
original scheme \cite{ABJ} was considered in our previous analyses
\cite{ewcorr,bsg-eps}, where
$\epsilon_{1,2,3}$ are defined from a basic set of observables $\Gamma_{l},
A^{l}_{FB}$ and $M_W/M_Z$. Because of the large $m_t$-dependent vertex
corrections to $\Gamma_b$, the $\epsilon_{1,2,3}$ parameters   and $\Gamma_b$
can be correlated only for a fixed value of $m_t$. Therefore, $\Gamma_{tot}$,
$\Gamma_{hadron}$ and $\Gamma_b$ were not included in Ref.~\cite{ABJ}. However,
in the new $\epsilon$-scheme, introduced recently in Ref.~\cite{ABC}, the above
difficulties are overcome by introducing a new parameter, $\epsilon_b$, to
encode the $\Zbb$ vertex corrections. The four $\epsilon$'s are now defined
from an enlarged set of $\Gamma_{l}$, $\Gamma_{b}$, $A^{l}_{FB}$ and $M_W/M_Z$
without even specifying $m_t$. Here we use this new $\epsilon$-scheme.
Experimentally, including all of the latest LEP data (complete 1992 LEP data
plus preliminary 1993 LEP data) allows one to determine most accurately the
allowed ranges for these parameters \cite{Altarelli}
\beq
\epsilon^{exp}_1=(1.8\pm3.1)\times10^{-3},\qquad
\epsilon^{exp}_b=(-0.5\pm5.1)\times10^{-3}\ . \label{exp_epsilons}
\eeq
We only discuss $\epsilon_1,\epsilon_b$ since only these parameters provide
constraints in supersymmetric models at the 90\%CL \cite{ewcorr,ABCII}.

The expression for $\epsilon_1$ is given by \cite{BFC}
\beq
\epsilon_1=e_1-e_5-{\delta G_{V,B}\over G}-4\delta g_A,\label{eps1}
\eeq
where $e_{1,5}$ are the following combinations of vacuum polarization
amplitudes
\begin{eqnarray}
e_1&=&{\alpha\over 4\pi \sin^2\theta_W M^2_W}[\Pi^{33}_T(0)-\Pi^{11}_T(0)],
\label{e1}\\
e_5&=& M_Z^2F^\prime_{ZZ}(M_Z^2),\label{e5}
\end{eqnarray}
and the $q^2\not=0$ contributions $F_{ij}(q^2)$ are defined by
\beq
\Pi^{ij}_T(q^2)=\Pi^{ij}_T(0)+q^2F_{ij}(q^2).
\eeq
The $\delta g_A$ in Eqn.~(\ref{eps1}) is the contribution to the axial-vector
form factor at $q^2=M^2_Z$ in the $Z\to l^+l^-$ vertex from proper vertex
diagrams and fermion self-energies, and $\delta G_{V,B}$ comes from the
one-loop box, vertex and fermion self-energy corrections to the $\mu$-decay
amplitude at zero external momentum. These non-oblique Standard Model
corrections are
non-negligible, and must be included in order to obtain an accurate Standard
Model
prediction.

The parameter $\epsb$ is defined from $\Gamma_b$, the inclusive partial width
for $\Zbb$, as follows \cite{ABC}
\begin{equation}
\Gamma_b=3 R_{QCD} {G_FM^3_Z\over 6\pi\sqrt 2}\left(
1+{\alpha\over 12\pi}\right)\left[ \beta _b{\left( 3-\beta
^2_b\right)\over 2}(g^b_V)^2+\beta^3_b (g^b_A)^2\right] \;,
\end{equation}
with
\begin{eqnarray}
R_{QCD} &\cong&\left[1+1.2{\alpha_S\left(
M_Z\right)\over\pi}-1.1{\left(\alpha_S\left(
M_Z\right)\over\pi\right)}^2-12.8{\left(\alpha_S\left(
M_Z\right)\over\pi\right)}^3\right] \;,\\
\beta_b&=&\sqrt {1-{4m_b^2\over M_Z^2}} \;, \\
g^b_A&=&-{1\over2}\left(1+{\epsilon_1\over2}\right)\left(
1+{\epsb}\right)\;,\\
{g^b_V\over{g^b_A}}&=&{{1-{4\over3}{\ov s}^2_W+\epsb}\over{1+\epsb}}\;.
\end{eqnarray}
where ${\ov s}^2_W$ is an effective $\sin^2\theta_W$ for on-shell $Z$.

In the calculation of $\epsilon_1$ we have included the complete supersymmetric
contributions to the oblique corrections. For increasing sparticle masses the
heavy sector of the theory decouples and only the Standard Model effects
survive (\ie, quadratic $m_t$-dependence of $\epsilon_{1,b}$ and logarithmic
$m_H$-dependence of $\epsilon_{1}$). On the other hand, relatively light
sparticle masses can give rise to significant deviations from the Standard
Model predictions. For $\epsilon_1$, a large negative shift can occur due to a
$Z$-wavefunction renormalization threshold effect through the $q^2$-dependence
in $e_5$ when $m_{\chi^\pm_1}\to{1\over2}M_Z$ \cite{BFC}.  The characteristic
features of supersymmetric contributions to $\epsilon_b$ are: (i) a negative
contribution from charged Higgs--top loops which grows as $m^2_t/\tan^{2}\beta$
for $\tan\beta\ll{m_t\over{m_b}}$; (ii) a positive contribution from
chargino-stop loops which in this case grows as $m^2_t/\sin^{2}\beta$; and
(iii) a contribution from neutralino(neutral Higgs)--bottom loops which grows
as $m^2_b\tan^{2}\beta$ and is negligible except for large values of
$\tan\beta$ (\ie, $\tan\beta\gsim{m_t\over{m_b}}$) (the contribution (iii) has
been neglected in our analysis).

Compared with the previous experimental values for the $\epsilon$ parameters
obtained by including the complete 1992 LEP data \cite{Altlecture} (which were
used in Ref.~\cite{eps1-epsb}) those in Eq.~(\ref{exp_epsilons}) have moved in
such a way that the Standard Model predictions have become in better agreement
with LEP data than before \cite{Altarelli,EFL}. In Fig.~\ref{Figure5} we
present the results of the calculation of $\epsilon_1$ and
$\epsilon_b$ (as described above) for all the allowed points in $SU(5)\times
U(1)$ supergravity in both moduli and dilaton scenarios, and for
$m_t=130,150,170,180\GeV$. In the figures we include three experimental
ellipses representing the 1-$\sigma$ (from Ref.~\cite{Altarelli}), 90\%CL, and
95\%CL experimental limits obtained from analyzing all of the latest LEP
electroweak data. The shift in the experimental data corresponds to a shift
in the center point of the ellipses towards larger values of $\epsilon_1$ and
smaller values of $\epsilon_b$. As a consequence, at the 90\%CL there are no
constraints from $\epsilon_b$ alone ({\em c.f.} Ref.~\cite{eps1-epsb}).
Nonetheless, the imposition of the correlated constraint (\ie, the ellipses),
is significantly more restrictive than imposing the $\epsilon_1$ constraint by
itself.

 For both scenarios, the effects of light charginos ($\chi^\pm_1$) and
stop-squarks ($\tilde t_{1,2}$), as described above, are rather pronounced.
At the 90\%CL there are no constraints for $m_t\lsim 170\GeV$, but for
$m_t=180\GeV$ only very light charginos ($m_{\chi^\pm_1}\lsim 70\GeV)$ are
allowed. Should the top quark be rather heavy, this light-chargino effect would
appear to be a sensible explanation. However, as we discuss below, other
experimental constraints on the parameter space for light supersymmetric
particles would make this possibility rather unlikely.

\section{Constraints on parameter space}
\label{constraints}
In this section we describe the experimental constraints which have
been applied to the points in the basic parameter space described in
Section \ref{model}. Each of these constraints leads to an excluded area
in the $(m_{\chi^\pm_1},\tan\beta)$ plane for a fixed value of $m_t$. Since
all sparticle masses scale with $m_{1/2}$, the lightest chargino mass is as
good a choice as any other one, and has the advantage of being readily
measurable. Our choices for $m_t$, \ie, $m_t=130,150,170,180\GeV$ are
motivated by the direct lower limit on the top-quark mass from Tevatron
searches ($m_t>131\GeV$ \cite{D0top}) and by the indirect estimates of
the mass from fits to the electroweak data ($m_t=140\pm20\GeV$
\cite{Altarelli,EFL}). The effect of each of the constraints is denoted
by a particular symbol on the parameter space plots in Figs. \ref{Figure6},
\ref{Figure7},\ref{Figure8},\ref{Figure9} for the various scenarios under
consideration. In all these figures there is an eye-guiding vertical dashed
line which corresponds to $m_{\chi^\pm_1}=100\GeV$. The purpose of
Fig.~\ref{Figure8} is to show where such line lies in the $(m_{\tilde
g},\tan\beta)$ plane. Kinematically speaking, the weakly interacting
sparticles (\ie, charginos) are more accessible than the strongly interacting
ones (\ie, gluino and squarks).

\subsection{$b\to s\gamma$}
\label{bsgamma}
The rare radiative flavor-changing-neutral-current (FCNC) $b\to s\gamma$ decay
has been observed by the CLEOII Collaboration in the following 95\% CL allowed
range \cite{Thorndike}
\beq
\bsg=(0.6-5.4)\times10^{-4}.\label{bsg}
\eeq
Since large enhancements and suppressions of $\bsg$, relative to the Standard
Model value, can occur in $SU(5)\times U(1)$ supergravity, the above allowed
interval can be quite restrictive \cite{bsgamma,bsg-eps} (see also
Ref.~\cite{BG,Oshimo}). The expression used for $\bsg$ is given by \cite{BG}
\beq
{B(b\to s\gamma)\over B(b\to ce\bar\nu)}={6\alpha\over\pi}
{\left[\eta^{16/23}A_\gamma
+\coeff{8}{3}(\eta^{14/23}-\eta^{16/23})A_g+C\right]^2\over
I(m_c/m_b)\left[1-\coeff{2}{3\pi}\alpha_s(m_b)f(m_c/m_b)\right]},
\label{brbsgeq}
\eeq
where $\eta=\alpha_s(M_Z)/\alpha_s(m_b)$, $I$ is the phase-space factor
$I(x)=1-8x^2+8x^6-x^8-24x^4\ln x$, and $f(m_c/m_b)=2.41$ is the QCD
correction factor for the semileptonic decay. In our analysis we use the
leading-order QCD corrections to the $b\to s\gamma$ amplitude when evaluated at
the $\mu=m_b$ scale \cite{GSW}, \ie, $C=\sum_{i=1}^8 b_i\eta^{d_i}=-0.1766$ for
$\eta=0.548$, with the $b_i,d_i$ coefficients given in Ref.~\cite{BG}.
In our computations we have used: $\alpha_s(M_Z)=0.118$,
$B(b\to ce\bar\nu)=10.7\%$, $m_b=4.8\GeV$, and $m_c/m_b=0.3$. The
$A_\gamma,A_g$ are the coefficients of the effective $bs\gamma$ and $bsg$
penguin operators evaluated at the scale $M_Z$. The contributions to $A_{\gamma
,g}$ from the $W-t$ loop, the $H^\pm-t$ loop, and the $\chi^\pm_i-\tilde t_k$
loop are given in Ref.~\cite{BG} in the justifiable limit of negligible gluino
and neutralino contributions and degenerate squarks (except for the $\tilde
t_{1,2}$) \cite{Bertolini}.

The results of the calculation in the moduli and dilaton scenarios are given in
Refs.~\cite{bsgamma,bsg-eps}. In both scenarios there exists a significant
region of parameter space where $\bsg$ is highly suppressed due to a phenomenon
involving a complicated cancellation against the QCD correction factor $C$
\cite{bsgamma,bsg-eps}. What happens is that in Eq.~(\ref{brbsgeq}), the
$A_\gamma$ term nearly cancels against the QCD correction factor $C$; the $A_g$
contribution is small.

The points in parameter space which are excluded at the 95\%CL are denoted by
pluses ($+$) in Figs.~\ref{Figure6} and \ref{Figure7} for the moduli and
dilaton scenarios respectively, and for the four chosen values of $m_t$.
The strict no-scale scenario (see Fig.~\ref{Figure9}a) is also constrained in
this fashion, although only for $m_t=130,150\GeV$. The special dilaton scenario
is not constrained by $\bsg$ (see Fig.~\ref{Figure9}b) because of the small
values of $\tan\beta$ required in this case. Note that the constraints are
generally much stricter for $\mu>0$.

\subsection{$(g-2)_\mu$}
\label{g-2}
The long-standing experimental value for the anomalous magnetic moment of the
muon $a_\mu$ for each sign of the muon electric charge \cite{oldg} can be
averaged to yield \cite{kinoII}
\beq
a^{exp}_\mu=1\ 165\ 923(8.5)\times10^{-9}.
\eeq
The uncertainty on the last digit is indicated in parenthesis. On the other
hand, the total standard model prediction is \cite{kinoII}
\beq
a^{SM}_\mu=1\ 165\ 919.20(1.76)\times10^{-9}. \label{g-2SM}
\eeq
Subtracting the experimental result gives \cite{kinoII}
\beq
a^{SM}_\mu-a^{exp}_\mu=-3.8(8.7)\times10^{-9},\label{diff}
\eeq
which is perfectly consistent with zero. The uncertainty in the experimental
determination of $a_\mu$ is expected to be reduced significantly (down to
$0.4\times10^{-9}$) by the new E821 Brookhaven experiment \cite{newg}, which is
scheduled to start taking data in late 1994. Any beyond-the-standard-model
contribution to $a_\mu$ (with presumably negligible uncertainty) will simply be
added to the central value in Eq.~(\ref{diff}). Therefore, we can obtain an
allowed interval for any supersymmetric contribution, such that
$a^{susy}_\mu+a^{SM}_\mu-a^{exp}_\mu$ is consistent with zero at the 95\%
confidence level,
\beq
-13.2\times10^{-9}<a^{susy}_\mu<20.8\times10^{-9}.\label{bounds}
\eeq

The supersymmetric contributions to $a_\mu$ in $SU(5)\times U(1)$ supergravity
have been recently computed in Ref.~\cite{g-2}. There it was
noted that a contribution to $a_\mu$, which is roughly proportional to
$\tan\beta$, leads to enhancements which can easily make $a^{susy}_\mu$ run in
conflict with the bounds given in Eq.~(\ref{bounds}). In general, there are two
sources of one-loop supersymmetric contributions to $a_\mu$: (i) with
neutralinos and smuons in the loop; and (ii) with charginos and sneutrinos in
the loop. The mixing angle of the smuon eigenstates is small and this
suppresses the neutralino-smuon contribution. Moreover, the various
neutralino-smuon contributions tend to largely cancel among themselves. This
implies that the chargino-sneutrino contributions are the dominant ones, and in
fact some of these are enhanced for large values of $\tan\beta$, as follows.

Picturing the chargino-sneutrino one-loop diagram, with the photon being
emitted off the chargino line, there are two ways in which the helicity of the
muon can be flipped, as is necessary to obtain a non-vanishing $a_\mu$:
\begin{description}
\item (i) It can be flipped by an explicit muon mass insertion on one of the
external muon lines, in which case the coupling at the vertices is between a
left-handed muon, a sneutrino, and the wino component of the chargino and has
magnitude $g_2$. It then follows that $a_\mu$ will be proportional to
$g^2_2(m_\mu/\tilde m)^2 |V_{j1}|^2$, where $\tilde m$ is a supersymmetric mass
in the loop and the $V_{j1}$ factor picks out the wino component of the $j$-th
chargino. This is the origin of the ``pure gauge" contribution to
$a^{susy}_\mu$.
\item (ii) Another possibility is to use the muon Yukawa coupling on one of
the vertices, which flips the helicity and couples to the Higgsino component
of the chargino. One also introduces a chargino mass insertion to switch to the
wino component and couple with strength $g_2$ at the other vertex. The
contribution is now proportional to $g_2\lambda_\mu(m_\mu m_{\chi^\pm_j}/\tilde
m^2)V_{j1}U_{j2}$, where $U_{j2}$ picks out the Higgsino component of the
$j$-th chargino. The muon Yukawa coupling is given by $\lambda_\mu=g_2
m_\mu/(\sqrt{2}M_W\cos\beta)$. This is the origin of the gauge-Yukawa
contribution to $a^{susy}_\mu$.
\end{description}
The ratio of the ``pure gauge" to the ``gauge-Yukawa" contributions is then
roughly
\beq
g^2_2\,(m_\mu/\tilde m)/(g_2\lambda_\mu)\sim g_2/\sqrt{1+\tan^2\beta},
\eeq
for $\tilde m\sim100\GeV$. Thus, for small $\tan\beta$ both contributions
are comparable, but for large $\tan\beta$ the ``gauge-Yukawa" contribution
is greatly enhanced.

The results of the calculation in the moduli and dilaton scenarios are given in
Ref.~\cite{g-2}. The points in parameter space which are excluded at the
95\%CL are denoted by crosses ($\times$) in Figs.~\ref{Figure6} and
\ref{Figure7} for the moduli and dilaton scenarios respectively, and for
the four chosen values of $m_t$. As expected, the $(g-2)_\mu$ constraint
has a similar effect for the two signs of $\mu$, and exclude the larger
values of $\tan\beta$ which are allowed for chargino masses up to about 100
GeV. The constraint appears less effective for $m_t=170,180\GeV$ (\ie, there
are fewer crosses), but this is just because for the larger values of $m_t$,
$\tan\beta$ is cut-off at smaller values. The strict no-scale scenario
(see Fig.~\ref{Figure9}a) is also constrained in this fashion, although only
for $m_t=130,150\GeV$. The special dilaton scenario is not constrained by
$(g-2)_\mu$ (see Fig.~\ref{Figure9}b) because of the small values of
$\tan\beta$ required in this case (\ie, $\tan\beta<1.64$).

\subsection{Neutrino telescopes}
\label{NT}
The basic idea is that the neutralinos ($\chi$) are assumed to make up the
dark matter in the galactic halo ---an important assumption which should
not be overlooked--- and can be gravitationally captured by
the Sun or Earth \cite{PS85,Gould}, after losing a substantial amount of
energy through elastic collisions with nuclei. The neutralinos captured in
the Sun or Earth cores annihilate into all possible ordinary particles,
and the cascade decays of these particles as well as their interactions with
the solar or terrestrial media produce high-energy neutrinos as
one of several end-products. These neutrinos can then travel from the
Sun or Earth cores to the vicinity of underground detectors, and interact
with the rock underneath producing detectable upwardly-moving muons.
Such detectors are called ``neutrino telescopes". The calculation of the
upwardly-moving muon fluxes induced by the neutrinos from the Sun
and Earth in $SU(5)\times U(1)$ supergravity has been performed in
Ref.~\cite{NT}. The currently most stringent 90\% C.L. experimental upper
bounds, obtained at Kamiokande, for neutrinos from the Sun \cite{KEKII} and
Earth \cite{KEK} are respectively
\begin{eqnarray}
\Gamma_{\rm Sun}&<&6.6\times10^{-14} {\rm cm}^{-2} {\rm s}^{-1}
=2.08\times10^{-2}{\rm m}^{-2}{\rm yr}^{-1},\label{eq:Sup}\\
\Gamma_{\rm Earth}&<&4.0\times10^{-14} {\rm cm}^{-2} {\rm s}^{-1}
=1.26\times10^{-2}{\rm m}^{-2}{\rm yr}^{-1}.\label{eq:Eup}
\end{eqnarray}

In order to calculate the expected rate of neutrino production due to
neutralino annihilation, it is necessary to first evaluate the rates at
which the neutralinos are captured in the Sun and Earth.
In our calculations, we follow a procedure similar to that of
Refs.~\cite{GGR91,KAM} in calculating the capture rate. To take into account
the effect of the actual halo neutralino relic density, we follow the
conservative approach of Ref.~\cite{GGR91} for the local neutralino density
$\rho_\chi$: (a) $\rho_\chi = \rho_h = 0.3\,{\rm GeV}/{\rm cm}^3$,
if $\Omega_\chi h^2_0 > 0.05$; while (b) $\rho_\chi = (\Omega_\chi
h^2_0/0.05)\rho_h$, if $\Omega_\chi h^2_0 \lsim 0.05$. Here $\rho_h$ is
the standard halo density. The dominant contribution to the capture cross
section is the coherent interaction due to the exchange of the two CP-even
Higgs bosons ($h$ and $H$), and the squarks. In addition, for capture by the
Sun, one also includes the spin-dependent cross section due to both $Z$-boson
exchange and squark exchange for the scattering from hydrogen. The computation
of the detection rate of upwardly-moving muons is rather involved and is
described in detail in Ref.~\cite{NT}.

For each point in the parameter spaces of the scenarios we consider, in
Ref.~\cite{NT} the relic abundance of neutralinos was determined and then the
capture rate was computed in the Sun and Earth, as well as the resulting
upwardly-moving muon detection rate. A particularly important feature of
the results is a kinematic enhancement of the capture rate by the Earth
because of the Fe nucleus. The capture and detection rates increase with
increasing $\tan\beta$, since the dominant piece of the coherent
neutralino-nucleon scattering cross section via the exchange of the lightest
Higgs boson $h$ is proportional to $(1+{\tan}^2\beta)$. Also, the capture rate
decreases with increasing $m_\chi$, since the scattering cross section
falls off as $m^{-4}_h$ and $m_h$ increases with increasing $m_\chi$.

The present experimental constraints from ``neutrino telescopes'' on the
parameter space are quite weak, as evidenced by the few excluded points
in Figs.~\ref{Figure6} and \ref{Figure7} (denoted by diamonds `$\diamond$').
In fact, the Kamiokande upper bound from the Earth capture is only useful to
exclude regions of the parameter space with $m_\chi \approx m_{\rm Fe}$ due to
the enhancement effect mentioned above. Because of the weakness of this
constraint, the effect has not been calculated for the special scenarios in
Fig.~\ref{Figure9}. Nonetheless, future improved sensitivity in underground
muon detection rates -- a factor of two with MACRO, a ten-fold improvement with
Super-Kamiokande, and factors of 20--100 from DUMAND and AMANDA -- should make
this constraint rather important, if neutralinos indeed constitute a
significant portion of the dark galactic halo.

\subsection{$\epsilon_1-\epsilon_b$}
\label{eps1b}
The updated calculation of $\epsilon_1,\epsilon_b$ has been given in Section
\ref{epsilons} above. The results are shown in Fig.~\ref{Figure5}. Here we
choose to constrain the parameter space by demanding theoretical predictions
which agree with experiment to better than 90\%CL, \ie, we exclude points in
parameter space which are outside the 90\%CL ellipses in Fig.~\ref{Figure5}.
This constraint entails restrictions for $m_t=180\GeV$ only.
Note that from our calculations, all $m_t=180\GeV$ points are allowed at the
95\%CL. However, more comprehensive analyses \cite{Altarelli,EFL} already
exclude $m_t=180\GeV$ at the 95\%CL (\ie, $m_t=140\pm20\GeV$), and thus our
restriction is in practice likely to be more statistically significant than can
be surmised from our analysis alone. The excluded points in parameter space are
shown as
squares `$\Box\,$' in Figs.~\ref{Figure6}d,\ref{Figure7}d, and \ref{Figure9}a.
The effect of this constraint is severe and, as discussed in
Section~\ref{epsilons}, requires rather light values of the chargino mass.
Moreover, such light values of $m_{\chi^\pm_1}$ are very likely to be
excluded by other constraints, as the figures show. This means that a
``light chargino effect" may not be a viable way out from a possible
experimentally heavy top quark.

\subsection{Trileptons}
\label{trileptons}
The process of interest is $p\bar p\to \chi^0_2\chi^\pm_1 X$, where both
neutralino and chargino decay leptonically: $\chi^0_2\to\chi^0_1 l^+l^-$, and
$\chi^\pm_1\to \chi^0_1 l^\pm\nu_l$, with $l=e,\mu$. The production cross
section proceeds through $s$-channel $W^*$-exchange and $t$-channel
squark-exchange (a small contribution). This signal, first studied in
Ref.~\cite{trileptons}, has been explored in $SU(5)\times U(1)$ supergravity
in Ref.~\cite{LNWZ}. The first experimental limits obtained by the D0
\cite{White} and CDF \cite{Kato,Kamon} Collaborations have been recently
announced. The irreducible backgrounds for this process are very
small, the dominant one being $p\bar p\to W^\pm Z\to
(l^\pm\nu_l)(\tau^+\tau^-)$ with a cross section into trileptons of
$(\sim1\pb)({2\over9})(0.033)(0.34)^2\sim1\fb$. Much larger ``instrumental"
backgrounds exist when for example in $p\bar p\to Z\gamma$, the photon
``converts" and fakes a lepton in the detector; with the present sensitivity,
suitable cuts have been designed to reduce this background to acceptable levels
\cite{White}.

The trilepton signal is larger in the moduli scenario because the
sleptons which mediate some of the decay channels can be on-shell and the
leptonic branching ratios are significantly enhanced (as large as $2\over3$)
relative to a situation with heavier sparticles in the dilaton scenario, where
the $W,Z$-exchange channels tend to dominate, and the leptonic branching
fractions are smaller \cite{LNWZ}. The results of these calculations have
been given in Ref.~\cite{LNWZ}. A distinctive feature of the $SU(5)\times U(1)$
supergravity predictions is that for light chargino masses the trilepton signal
can be rather small in the moduli scenario. This occurs when the neutralino
leptonic branching fraction is suppressed because the sneutrinos are on-shell
and the $\chi^0_2\to\nu\tilde\nu$ channel dominates.

The present experimental limits \cite{White,Kato,Kamon} from the Tevatron are
rather weak, with sensitivity for $m_{\chi^\pm_1}\lsim50\GeV$ only \cite{Kato}.
In the case of $SU(5)\times U(1)$ supergravity, no points in parameter space
are excluded by the present experimental limits. However, with the projected
increase in integrated luminosity during 1994, this experimental constraint
could soon become relevant, as we discuss in Section~\ref{Tevatron} below.

\subsection{Updated Higgs-boson mass limit}
\label{mh}
The current LEPI lower bound on the Standard Model (SM) Higgs boson mass
stands at $m_H>63.8\GeV$ \cite{Sopczak}. This bound is obtained by studying the
process $e^+e^-\to Z^*H$ with subsequent Higgs-boson decay into two jets. The
MSSM analog of this production process leads to a cross section differing just
by a factor of $\sin^2(\alpha-\beta)$. In Ref.~\cite{LNPWZh} it was shown that
in supergravity models with radiative electroweak symmetry breaking, as is
the case of $SU(5)\times U(1)$ supergravity, the lightest Higgs boson behaves
very much like the Standard Model Higgs boson. In particular, the
$\sin^2(\alpha-\beta)$ factor approaches unity as the supersymmetry mass scale
is raised. The branching fraction $B(h\to b\bar b)$ also approaches the
Standard Model value, although one has to watch out for new supersymmetric
decays, most notably $h\to\chi^0_1\chi^0_1$. In any event, a straightforward
procedure to adapt the experimental lower bound on the Standard Model
Higgs-boson mass to the supersymmetric case is described in Ref.~\cite{LNPWZh}.
The following condition must be satisfied for allowed points in parameter space
\cite{LNPWZh,LG}
\beq
f\cdot\sin^2(\alpha-\beta)<P(M_H^{min}/M_Z)/P(m_h/M_Z),
\eeq
where $M_H^{min}$ is the experimental lower bound on the Standard Model
Higgs-boson mass (\ie, $M_H^{min}=63.8\GeV$), and we have used the fact that
the cross sections differ only by the coupling factor $\sin^2(\alpha-\beta)$
and the Higgs-boson mass dependence, which enters through a function $P$
\cite{HHG}
\beq
P(y)={3y(y^4-8y^2+20)\over\sqrt{4-y^2}}\cos^{-1}
\left(y(3-y^2)\over2\right)-3(y^4-6y^2+4)\ln
y-\coeff{1}{2}(1-y^2)(2y^4-13y+47).
\eeq

The determination of the basic parameter space in Section~\ref{model}, includes
the LEP experimental limits on sparticle masses and the experimental limit
$M_H^{min}=61.3\GeV$. The updated experimental limit of $M_H^{min}=63.8\GeV$
excludes some further points in parameter space (denoted by octagon symbols
in Figs.~\ref{Figure6},\ref{Figure7},\ref{Figure9}) for the smallest values
of $\tan\beta$ and $m_t=130,150\GeV$.

\section{The allowed parameter space}
\label{allowed}
The constrained parameter spaces shown in Figs.~\ref{Figure6},\ref{Figure7},
\ref{Figure9} show some regularities which are worth pointing out. First,
the constraints for $\mu<0$ are generally weaker than those for $\mu>0$
because the $b\to s\gamma$ constraint is basically inoperative for
$\mu<0$. It is also clear that the region to the left of the dashed line
($m_{\chi^\pm_1}<100\GeV$) is rather restricted. This region represents the
area of sensitivity at LEPII from direct chargino searches. (LEPII could
greatly extend this region through Higgs-boson searches though.)

For $m_t=180\GeV$ things are very constrained. The most important constraint
comes from the $\epsilon_1-\epsilon_b$ ellipses in Fig.~\ref{Figure5}.
Moreover, the remaining allowed points, which require rather light chargino
masses ($m_{\chi^\pm_1}\lsim70\GeV$), are quite often in conflict with
other experimental constraints. The few remaining points in parameter space
have $\tan\beta\lsim8\,(12)$ in the moduli (dilaton) scenario. Also,
$m_{\chi^\pm_1}\lsim68\,(66)\GeV$ and $m_{\chi^\pm_1}\lsim65\,(68)\GeV$
for $\mu>0\,(\mu<0)$ in the moduli and dilaton scenarios respectively. For both
scenarios, a more sensitive measurement of the $b\to s\gamma$ branching
fraction is likely to probe the remaining allowed points for $m_t=180\GeV$.
Also, for $\mu<0$ in both scenarios, the expected increased sensitivity in
trilepton searches is likely to probe about half of the remaining points.

It is interesting to wonder if the present experimental constraints show any
preference for particular values of the top-quark mass. To explore this
question we carry out the following exercise: we count the number of points
in parameter space which are allowed for a fixed value of $m_t$. We do this
in two steps (see Fig.~\ref{Figure10}): (i) first imposing only the basic
theoretical and LEP experimental constraints (``theory+LEP") and (ii) imposing
in addition all of the experimental constraints described in
Section~\ref{constraints} (``ALL"). The result in Fig.~\ref{Figure10} is
interesting. The drop in the ``theory+LEP" curves near $m_t=190\GeV$ has been
studied in detail (for $m_t=180,185,187,188,189\GeV$) and corresponds to
encountering a Landau pole in the top-quark Yukawa coupling below the string
scale \cite{DL}. The ``ALL" curves show some $m_t$-dependence, although at
the moment no marked preference for particular values of $m_t$ is apparent
(besides the requirement of $m_t\lsim180\GeV$). Note that in spite of the
intricate dependence of the sparticle and Higgs-boson masses on the top-quark
mass (through the running of the RGEs and the radiative breaking mechanism),
the overall size of the parameter space does not depend so critically on $m_t$.

One can repeat the above exercise to see if any trends on the preferred
value of $\tan\beta$ appear. This time we count the number of allowed points
in parameter space for a given value of $\tan\beta$, for fixed $m_t$. The
resulting distribution is shown in Fig.~\ref{Figure11} for $m_t=150\GeV$.
Qualitatively similar distributions are obtained for other values of $m_t$.
In this case we discover that for $\mu>0$ there is a significant preference
towards the smaller values of $\tan\beta$. This result is apparent from
Figs.~\ref{Figure6},\ref{Figure7} also, and is mostly a consequence of the
$b\to s\gamma$ constraint, which is only efficient for $\mu>0$.

Future improvements in sensitivity on the experimental constraints which we
have imposed here, or the advent of new experimental constraints, may sharpen
the ``preditions" for the preferred values of $m_t$ and $\tan\beta$ obtained
in this statistical exercise.

\section{Prospects for direct experimental detection}
\label{direct}
In this section we consider the still-allowed parameter space, \ie, the
points marked by dots in Figs.~\ref{Figure6},\ref{Figure7},\ref{Figure9},
and the prospects for their direct experimental detection. In this section
we consider only the representative value of $m_t=150\GeV$.

\subsection{Tevatron}
\label{Tevatron}
The Tevatron can explore the supersymmetric spectrum through the traditional
missing energy signature in the decay of the strongly interacting
gluinos and squarks, or through the trilepton signal in the decay of the
weakly interacting charginos and neutralinos. Here we concentrate on the
latter signal, whose calculation has been described in
Section~\ref{trileptons}. The cross section $\sigma(p\bar p\to
\chi^\pm_1\chi^0_2X)$ (for $\sqrt{s}=1.8\TeV$) is shown in Fig.~\ref{Figure12}
for $m_t=150\GeV$ in the moduli and dilaton scenarios (top row), and shows
little variation from one scenario to the other. Moreover, the results for
other values of $m_t$ are qualitatively the same and quantitatively quite
similar. On the bottom row of Fig.~\ref{Figure12} we show the cross section
into trileptons, \ie, with the leptonic branching fractions included. The
95\%CL experimental upper limit from CDF is also indicated.\footnote{Note that
the experimental numbers in Ref.~\cite{Kato} apply to a single channel
(\ie, $eee$, $ee\mu$, $e\mu\mu$, or $\mu\mu\mu$), and need to be multiplied
by four to be compared with our predictions for the total $e+\mu$ trilepton
rate.} As mentioned above, in the moduli
scenario the neutralino leptonic branching fraction can be suppressed for light
chargino masses. Note that in the dilaton scenario such suppression is not
manifest because of the heavier sparticle mass spectrum. In Fig.~\ref{Figure13}
we show the analogous results for the strict no-scale and special dilaton
scenarios, neither of which show a suppression for light chargino masses.

The present experimental limits from CDF have been obtained by analyzing
approximately $18\ipb$ of data. By the end of the 1994 run it is expected that
each detector will be delivered $75\ipb$, of which CDF should be able to
collect say 80\%. Therefore, CDF could expect to have about $80\ipb$ by the end
of the run, which is 4 times the present amount of data. A similar situation is
expected from D0. Moreover, the center-of-mass energy will be increased to
nearly 2 TeV, which implies a $\approx30\%$ increase in the
chargino-neutralino cross section for 100 GeV charginos. Since tougher cuts
will be required to suppress the backgrounds with the increased sensitivity, as
a working estimate we have assumed that the new limits (if no signal is
observed) will be down by a factor of four from those shown on
Fig.~\ref{Figure12}. We are then able to identify points in the still-allowed
parameter space which could be probed by the end of 1994. (The Tevatron will
likely not run again until 1997.) These are shown as pluses ($+$) in
Fig.~\ref{Figure14} for the moduli and dilaton scenarios. Note that with the
increased sensitivity, chargino masses as high as $\approx100\GeV$ could be
probed in the moduli scenario. The rates are smaller in the dilaton scenario.
Note also that this probe is much more sensitive for $\mu<0$ (see
Fig.~\ref{Figure12}).

Searches for squarks and gluinos at the Tevatron are at kinematical
disadvantage in the model under consideration. Indeed, compare the relative
position of the dashed vertical line on Fig.~\ref{Figure14}, with the
corresponding line on Fig.~\ref{Figure8}. The near-future trilepton searches
correspond mostly to gluino and squark masses in the range $(300-400)\GeV$,
which are probably beyond the direct reach of the Tevatron for the same data
set.

\subsection{LEPII}
\subsubsection{Lightest Higgs boson}
Perhaps the single most useful piece of information that could come out of
LEPII is a measurement of the lightest Higgs-boson mass. Moreover, if the
Higgs boson is not observed at LEPII, because of limited statistics or
kinematics, still a strong constraint will follow for a large class of
supersymmetric models, in particular the ones under consideration here.
In Fig.~\ref{Figure15} we show the lightest Higgs-boson mass versus $\tan\beta$
for $m_t=150\GeV$ in the moduli and dilaton scenarios. Along each vertical
line the chargino mass increases from bottom to top. The dotted portions of
the lines are already excluded by the various constraints discussed in
Section~\ref{constraints}. For $m_t=150\GeV$ we find $m_h<118\GeV$. In
Fig.~\ref{Figure16} we consider the strict no-scale and special dilaton
scenarios. Since in these cases the value of $\tan\beta$ is determined (see
Fig.~\ref{Figure9}), the plot is against the chargino mass. In both
Fig.~\ref{Figure15} and \ref{Figure16} the horizontal line indicates the
limit of sensitivity of LEPII for $\sqrt{s}=200\GeV$, as we shortly discuss.
First let us note, as pointed out in Ref.~\cite{LNZII}, that the special
dilaton scenario (see Fig.~\ref{Figure16}) should be completely explored
at LEPII (even with $\sqrt{s}=190\GeV$) since $m_t\lsim155\GeV$ is required
in this case (see Section~\ref{specialdilaton}).

In $SU(5)\times U(1)$ supergravity, the dominant Higgs-boson production
mechanism at LEPII is $e^+e^-\to Z^*\to Zh$. This cross section differs from
its Standard Model counterpart only by a factor of $\sin^2(\alpha-\beta)$. Here
we find that generally $\sin^2(\alpha-\beta)>0.96$, in agreement with a general
result to this effect \cite{LNPWZh}.\footnote{In $SU(5)\times U(1)$
supergravity, the $e^+e^-\to hA$ channel is rarely kinematically allowed at
LEPII (since $m_A>m_h$) and is further suppressed by the small values of
$\cos^2(\alpha-\beta)$.} The usual analysis of the $b$-tagged Higgs-boson
signal at LEPII also requires the $h\to b\bar b$ branching fraction. If we
define
$f\equiv B(h\to b\bar b)/B(H_{SM}\to b\bar b)$, then the expected limit of
sensitivity at LEPII, $\sigma(e^+e^-\to Z^*\to ZH_{SM})>0.2\pb$ \cite{Sopczak}
becomes
\beq
\sigma(e^+e^-\to Z^*\to Zh)\times f>0.2\pb.\label{hlimit}
\eeq
This size signal is needed to observe a $3\sigma$ effect over background with
${\cal L}=500\ipb$.  Our results for this quantity are shown in
Fig.~\ref{Figure17}, along with the sensitivity limit in Eq.~(\ref{hlimit}).
Note that most points in parameter space accumulate along a well defined line.
This line corresponds to the Standard Model result. (Deviations from the line
are discussed below.) For $\sqrt{s}=200\GeV$, the limit of sensitivity in
Eq.~(\ref{hlimit}) translates into $m_h\lsim105\GeV$, while for
$\sqrt{s}=210\GeV$, $m_h\lsim115\GeV$ is obtained. From Fig.~\ref{Figure17} it
would still appear that even with $\sqrt{s}=210\GeV$, some points in parameter
space for $m_t=150\GeV$ would remain unreachable. However, detailed studies
\cite{Sopczak} show that for Higgs-boson masses away from the $Z$-pole (here we
are interested in $m_h\approx115-118\GeV$) the limit of sensitivity could be
improved to $(0.05-0.15)\pb$, and thus the whole parameter space for
$m_t=150\GeV$ and all scenarios considered could be explored at LEPII with
$\sqrt{s}=210\GeV$. Note that the same conclusion is obtained for
$m_t<150\GeV$, since the values of $m_h$ are lower then (\eg, for $m_t=130\GeV$
we find $m_h\lsim105\GeV$), and even smaller luminosities or beam energies may
suffice.

In Fig.~\ref{Figure17} there are some points which ``fall off" the main
curve. These correspond to suppressed values of the $h\to b\bar b$ branching
fraction (\ie, $f<1$) which occur when the invisible supersymmetric decay
channel $h\to\chi^0_1\chi^0_1$ is kinematically open \cite{LNPWZ}, as shown in
Fig.~\ref{Figure18}. However, the fraction of points in parameter space where
this happens is rather small (less than 10\%). Nonetheless, most of these
special points are still within the limit of sensitivity in Eq.~(\ref{hlimit})
and should not escape detection. The scarcity of points in parameter space
where the Higgs boson could decay invisibly may discourage detailed studies of
such signature in $SU(5)\times U(1)$ supergravity. However, when the invisible
mode is allowed, its branching fraction can be as large as 60\%.

We can see the effect on the parameter space of a possible measurement of $m_h$
by studying the Higgs-boson mass contours shown in Fig.~\ref{Figure14}, or
for the full parameter space in Fig.~\ref{Figure19new}. In general one would
obtain a constraint giving $\tan\beta$ for a given chargino mass. Moreover, a
minimum value of the chargino mass would be required, if $m_h\gsim100\GeV$.
Furthermore, in the strict no-scale and special dilaton scenarios, the chargino
mass itself would be determined (see Fig.~\ref{Figure16}) and thus the whole
spectrum. If only a lower bound on $m_h$ is obtained, still large portions of
the parameter space could be excluded, \ie, all of the areas to the left of the
corresponding mass contour.

What if $m_t>150\GeV$? For $m_t=170\GeV$, one obtains $m_h\lsim128\GeV$ and
$\sqrt{s}=240\GeV$ would be required for a full exploration of the parameter
space at LEPII.

We close this section with a last-minute remark. Two-loop QCD corrections to
$m_h$ have been recently shown to decrease the Higgs-boson mass by a
non-negligible amount \cite{HH}. A complete calculation of this effect in
$SU(5)\times U(1)$ supergravity is beyond the scope of this paper. A rough
assessment of the effects indicates that the Higgs-boson mass contours in
Fig.~\ref{Figure19new} (see also Fig.~\ref{Figure14}) would likely shift to
lower values.
%: $114\to108$, $112\to107$, $110\to105$, $105\to102$, $100\to98$.
This downward shift implies an enlarged reach for LEPII. Equivalently, the
above conclusions would require even lower values of the center-of-mass energy
or integrated luminosity.

\subsubsection{Charginos}
The cross section for chargino pair production is the largest of
all cross sections involving charginos and neutralinos at LEPII. In the
context of $SU(5)\times U(1)$ supergravity this has been shown in
Ref.~\cite{LNPWZ}. The most studied signature is the so-called mixed mode,
where one chargino decays leptonically and the other one hadronically. If the
chargino decay channels are dominated by $W$-exchange (\ie, branching ratio
into electron+muon is 2/9 and branching ratio into jets is 2/3) then the mixed
channel has a rate six times larger than the dilepton channel. The mixed
signature still has to contend with the $W^+W^-$ background. However, a series
of cuts have been designed which take advantage of different values for the
misssing mass, the mass of the hadronic system, and the mass of the
lepton+neutrino system when one considers the background and the signal
separately \cite{Grivaz}. In the case of $SU(5)\times U(1)$ supergravity,
$W$-exchange is not expected to dominate in chargino decay \cite{LNWZ}. In
fact, in the moduli scenario the sleptons are lighter and can therefore be
on-shell, thus enhancing the leptonic branching fraction to its maximum value
of 2/3. When this occurs the mixed signal is negligible, because of the much
suppressed hadronic branching fraction. In Fig.~\ref{Figure19} we show the
cross section for the mixed signal at LEPII for $\sqrt{s}=200\GeV$ and
$m_t=150\GeV$, for both moduli and dilaton scenarios. As expected, the mixed
rate is small (even vanishing for $\mu<0$!) in the moduli scenario. In the
dilaton scenario the rate is larger, but still much smaller than the
corresponding rate in a model where $W$-exchange dominates chargino decays.
This situation in fact occurs in the minimal $SU(5)$ supergravity model where
the rate is typically in the range
of $(1.5-2)\pb$, as shown in Fig. 3 of Ref.~\cite{LNWZ}. The signal is further
suppressed in the dilaton scenario because of a negative interference effect
between the $t$-channel sneutrino-exchange and the $s$-channel $\gamma^*-$ and
$Z^*$-exchange \cite{LNPWZ}. Despite all these suppression factors, the mixed
signal is still quite observable, as we now discuss.

The various cuts on the $W^+W^-$ background mentioned above manage to
suppress it down to 9fb \cite{Grivaz}, while the signal (assuming $W$-exchange
dominance) is suppressed by a factor of about $\epsilon=0.4$. Assuming that
$\epsilon$ is not too different in our case, to observe a $5\sigma$ effect one
would require:
\beq
{\epsilon(\sigma B)_{mixed}\,{\cal L}\over\sqrt{0.009{\cal L}}}>5\Rightarrow
(\sigma B)_{mixed}>{1.18\over\sqrt{{\cal L}}}=
\left\{
\begin{array}{ll}
0.12\pb,&{\cal L}=100\ipb\\
0.05\pb,&{\cal L}=500\ipb
\end{array}
\right.\,.\label{mixedlimit}
\eeq
The sensitivity limit obtained in this way for ${\cal L}=500\ipb$ is shown
as a horizontal dashed line on Fig.~\ref{Figure19}. The points in parameter
space which would be probed in this way are marked by crosses ($\times$) in
Fig.~\ref{Figure14}. In the dilaton scenario one could thus probe nearly all
points up to the kinematical limit (\ie, $m_{\chi^\pm_1}<100\GeV$).

Before concluding this section, let us examine the dilepton mode in chargino
pair production, since in the moduli scenario it is likely to have a much
larger rate than the mixed mode does. The dilepton rate is shown in
Fig.~\ref{Figure20} for both scenarios. The real problem here is the taming of
the irreducible dilepton background from $W^+W^-$ production, \ie,
$\sigma(e^+e^-\to W^+W^-\to l^+\nu_l l^-\bar\nu_l)\approx(18)({2\over9})
({2\over9})=0.9\pb$ at $\sqrt{s}=200\GeV$. Cuts are apparently not very
efficient in suppressing this background \cite{Dionisi}, although a
re-assessment of this problem needs to be performed to be certain. In any
event, demanding that the dilepton signal have a $5\sigma$ significance over
this background implies
\beq
{(\sigma B)_{dilepton}\,{\cal L}\over\sqrt{0.9{\cal L}}}>5
\Rightarrow (\sigma B)_{dilepton}>
\left\{
\begin{array}{ll}
0.47\pb,&{\cal L}=100\ipb\\
0.21\pb,&{\cal L}=500\ipb
\end{array}
\right.\,.\label{dileptonlimit}
\eeq
This sensitivity limit is shown as a dashed line in Fig.~\ref{Figure20}. The
regions of parameter space possibly explorable in this way are not shown in
Fig.~\ref{Figure14} since the di-electron signal from selectron pair production
(discussed next) is much larger. Moreover, only 25\% of the dilepton signal
from chargino pair production consists of di-electrons.

\subsubsection{Sleptons}
The charged sleptons ($\tilde e_{L,R},\tilde\mu_{L,R},\tilde\tau_{L,R}$)
could be pair-produced at LEPII if light enough, and offer an interesting
supersymmetric signal through the dilepton decay mode. In the moduli
scenario there is a significant portion of the parameter space where these
particles are kinematically accessible at LEPII, while in the dilaton scenario
the accessible region is very small and will be neglected in what follows.
The cross sections of interest are
\begin{eqnarray}
e^+e^-&\to&\tilde e^+_L\tilde e^-_L,\tilde e^+_R\tilde e^-_R,
\tilde e^\pm_L\tilde e^\mp_R, \\
e^+e^-&\to&\tilde \mu^+_L\tilde \mu^-_L,\tilde \mu^+_R\tilde \mu^-_R,\\
e^+e^-&\to&\tilde \tau^+_L\tilde \tau^-_L,\tilde \tau^+_R\tilde \tau^-_R.
\end{eqnarray}
The $\tilde e^+_L\tilde e^-_L,\tilde e^+_R\tilde e^-_R$ final states receive
contributions from $s$-channel $\gamma^*$ and $Z^*$ exchanges and $t$-channel
$\chi^0_i$ exchanges, while the $\tilde e^\pm_L\tilde e^\mp_R$ only proceeds
through the $t$-channel. The $\tilde \mu^+_L\tilde \mu^-_L,\tilde \mu^+_R\tilde
\mu^-_R$ and $\tilde \tau^+_L\tilde \tau^-_L,\tilde \tau^+_R\tilde \tau^-_R$
final states receive only $s$-channel contributions, since all couplings are
lepton flavor conserving, and therefore mixed LR final states are not allowed
for smuon or stau production. In Fig.~\ref{Figure21} we show the total
selectron and total smuon cross sections, which include all the kinematically
accessible final states mentioned above. The results for stau pair-production
are very similar to those for smuon pair-production. The horizontal line
represents an estimate of the limit of sensitivity achievable with ${\cal
L}=500\ipb$, as given in Eq.~(\ref{dileptonlimit}) to observe a $5\sigma$
signal over the irreducible $W^+W^-$ dilepton background. The selectron
cross section is considerably larger than the smuon one because of the
additional production channels.

Our discussion in effect assumes that the acoplanar dilepton signal associated
with selectron pair production comes entirely from $\tilde e^\pm_{L,R}\to
e^\pm\chi^0_1$ decay channels, \ie, purely di-electrons, and similarly for
the smuon case. This is an approximation which holds fairly well in the
moduli scenario \cite{LNPWZ}.

The points in parameter space in the moduli scenario which would be
explorable through selectron searches at LEPII are shown in Fig.~\ref{Figure14}
as diamonds ($\diamond$). The corresponding points explorable through smuon
searches are not shown since the signal is smaller than in the selectron
case. A rather interesting result is that the indirect reach in the chargino
mass can be extended beyond the direct reach (of about 100 GeV). This effect
depends on the value of $\tan\beta$, and is relevant only for $\tan\beta\lsim6$
and $\mu>0$, as Fig.~\ref{Figure14} shows. In fact, the three dotted lines
for $\mu>0$ in Fig.~\ref{Figure21} correspond from left to right to
$\tan\beta=6,4,2$ respectively.

\subsection{HERA}
The weakly interacting sparticles may be detectable at HERA in $SU(5)\times
U(1)$ supergravity \cite{hera}. However, the mass range accessible is rather
limited, with only the moduli scenario being partially reachable. The
elastic scattering signal, \ie, when the proton remains intact, is the most
promising one. The deep-inelastic signal has smaller rates and is plagued
with large backgrounds \cite{hera}. The reactions of interest are
$e^- p\to \tilde e^-_{L,R}\chi^0_{1,2}p$ and $e^-p\to \tilde\nu_e\chi^-_1p$.
The total elastic supersymmetric signal is shown in Fig.~\ref{Figure22}
versus the chargino mass. The dashed lines represent limits of sensitivity
with ${\cal L}=100\ipb$ and $1000\ipb$ which will yield five ``supersymmetric"
events. This is a rather small signal. Moreover, considering the timetable for
the LEPII and HERA programs, it is quite likely that LEPII would explore all
of the HERA accessible parameter space before HERA does. This outlook may
change if new developments in the HERA program would give priority to the
search for the right-handed selectron ($\tilde e_R$) which could be rather
light in the moduli scenario of $SU(5)\times U(1)$ supergravity.

\section{Prospects for indirect experimental detection}
\label{indirect}
In section~\ref{constraints} we discussed four indirect
(\ie, $B(b\to s\gamma)$, $(g-2)_\mu$, neutrino telescopes, and
$\epsilon_1-\epsilon_b$) and two direct (\ie, trileptons and the lightest
Higgs-boson mass at LEPI) experimental constraints on the parameter space
of $SU(5)\times U(1)$ supergravity. Of the indirect constraints, the
neutrino telescopes probe may become strict in the not-so-distant future
(\ie, when MACRO comes into operation), however the implicit assumption of
significant neutralino population in the galactic halo cannot be verified
directly, and this diminishes the weight to be assigned to this constraint.
The $\epsilon_1-\epsilon_b$ constraint on the top-quark mass should become
stricter with the reduction of the present error bars by a factor of two
by the end of the LEPI program. In this section we examine the two remaining
indirect constraints ($B(b\to s\gamma)$ and $(g-2)_\mu$) for the still-allowed
points in parameter space.

In Fig.~\ref{Figure23} we show the values of $B(b\to s\gamma)$ calculated
for the still-allowed points in parameter space (for $m_t=150\GeV$) in
the moduli and dilaton scenarios. For reference, the whole range of possible
values before the imposition of the constraints discussed in
section~\ref{constraints} is addressed in Refs.~\cite{bsgamma,bsg-eps}. In the
moduli case, for $\mu>0$ one obtains a set of orderly lines for the indicated
values of $\tan\beta$, which keep increasing in steps of two beyond the values
explicitly noted. These lines in effect reach a minimum value of zero and
would have risen again, but this happens for excluded points in parameter
space. The various points not following along the orderly lines are remnants of
this behavior. For $\mu<0$ the results show little variability, spanning the
range $(3.4-4.0)\times10^{-4}$. In the dilaton scenario the qualitative
picture is somewhat similar, but for $\mu<0$ there is a much wider range
of possible values. For comparison, in the Standard Model for $m_t=150\GeV$
one gets $B(b\to s\gamma)_{SM}\approx4\times10^{-4}$ (although QCD corrections
need to be accounted for carefully). A more precise measurement of this
branching fraction should be used to exclude points in parameter space which
deviate significantly from the Standard Model prediction. A detailed
calculation of the QCD corrections in the supersymmetric case would be required
to make a careful comparison with the Standard Model predictions.

In Fig.~\ref{Figure24} we show the values of $a^{susy}_\mu$ versus the gluino
mass for $m_t=150\GeV$ in the moduli and dilaton scenarios. Reference values
of $\tan\beta$ are indicated. The dotted portions of the lines correspond to
points in parameter space excluded by the combined constraints in
section~\ref{constraints}. Note that for $\mu>0$ in both scenarios there is
a range of $a^{susy}_\mu$ values which is excluded for all values of
$\tan\beta$. The new Brookhaven E821 experiment is expected to achieve a
precision of $0.4\times10^{-9}$, which would entail a determination of
$\tan\beta$ as a function of the gluino (or chargino) mass. We remark that
the supersymmetric contributions to $a_\mu$ could be so large that the
uncertainty in the Standard Model prediction ($1.76\times 10^{-9}$, see
Eq.~(\ref{g-2SM})) would be basically irrelevant when testing a large fraction
of the allowed parameter space.

\section{Conclusions}
\label{conclusions}
We have presented an analysis of the several direct and indirect experimental
constraints which exist at present on the parameter space of $SU(5)\times U(1)$
supergravity in the moduli and dilaton scenarios and their special cases
(strict no-scale and special dilaton). These scenarios are inspired by possible
model-independent supersymmetry breaking scenarios in string models, and have
the non-automatic virtue of implying universal soft-supersymmetry-breaking
parameters. The scenarios can be described in terms of three parameters
$(m_{\chi^\pm_1},\tan\beta,m_t)$ which will be reduced down to two once the
top-quark mass is measured. This minimality of parameters is very useful in
correlating the many experimental predictions and constraints on the model.
The $(m_{\chi^\pm_1},\tan\beta)$ plane (for fixed $m_t$) has been discretized
and each point scrutinized to determine if the theoretical and basic LEP
experimental constraints are satisfied. For satisfactory points we have then
computed $B(b\to s\gamma)$, $(g-2)_\mu$, the rate of underground muon fluxes,
$\epsilon_1-\epsilon_b$, and the trilepton rate at the Tevatron. Generally
we find $m_t\lsim180\GeV$ to satisfy the $\epsilon_1-\epsilon_b$ constraint,
and some excluded regions of parameter space for specific values of $m_t$.

For the still-allowed points in parameter space we have re-evaluated the
experimental situation at the Tevatron, LEPII, and HERA. We have delineated the
region of parameter space that would be explored in the 1994 Tevatron run,
and by Higgs-boson, slepton, and chargino searches at LEPII with ${\cal
L}=500\ipb$. With estimates for the possible sensitivities at these colliders,
we conclude that the Tevatron could explore the parameter space with chargino
masses as high as $100\GeV$. On the other hand, searches for the lightest Higgs
boson at LEPII could explore all of the allowed parameter space in both
scenarios if $m_t\lsim150\GeV$ and the beam energy is raised up to
$\sqrt{s}=210\GeV$ (or lower if the two-loop QCD corrections to $m_h$ are
accounted for). In fact, a measurement of the Higgs-boson mass in the Standard
Model will almost uniquely determine the mass of the lightest Higgs-boson in
$SU(5)\times U(1)$ supergravity, since the relevant cross section and branching
fractions deviate little from their Standard Model counterparts. Because of the
mass correlations in the model, searches for selectrons allow LEPII to reach
into the parameter space beyond the direct reach for chargino masses (\ie,
$m_{\chi^\pm_1}<100\GeV$), thus selectrons
are the next-deepest probe of the parameter space (after the Higgs boson),
and charginos are the third probe. Searches for sparticles at HERA are not
competitive with those at LEPII, although supersymmetric particles in the
moduli scenario (in particular the right-handed selectron $\tilde e_R$) may be
light enough to be eventually observed at HERA. Searches for strongly
interacting sparticles (squarks and gluinos) are not kinematically favored at
the Tevatron since for example, chargino masses of 100 GeV correspond to gluino
and squark masses around 400 GeV. All of these possible constraints from future
direct particle searches have been shown in plots of the still-allowed points
in parameter space (see Fig.~\ref{Figure14}). These plots show the regions
where the various searches are sensitive and should serve as a `clearing house'
where the many experimental constraints are brought in, enforced, and their
implications discussed.

Let us conclude with a few general remarks in the context of $SU(5)\times U(1)$
supergravity (see Fig.~\ref{Figure14}):
\begin{itemize}
\item  If the Tevatron sees sparticles (charginos), then almost
certainly would LEPII see sparticles too.
\item  If the Tevatron does not see sparticles (charginos), not much
can be said about the prospects at LEPII.
\item  It is quite possible that LEPII would see the lightest Higgs boson but
no sparticles, if the Higgs-boson mass exceeds some $m_t$-dependent limit
($m_h\gsim105\GeV$ for $m_t=150\GeV$).
\item It is unlikely, although possible that LEPII would see sparticles but no
Higgs boson.
\item If LEPII sees the lightest Higgs boson, then we would get a line in the
$(m_{\chi^\pm_1},\tan\beta)$ plane, \ie, $\tan\beta$ as a function of
$m_{\chi^\pm_1}$ (for fixed or known $m_t$). The measurement would be
conclusive by itself only in the strict moduli and special dilaton scenarios.
\item  If the Higgs boson, and selectrons or charginos are seen
at LEPII, this should be enough to test the model decisively because of the
predicted correlations among the various predictions.
\end{itemize}

In sum, the analytical procedure proposed in this paper could be applied to
any supergravity model, and would serve as a standard against which
the feasibility of various models could be measured and compared.

%\vfill\eject
\section*{Acknowledgements}
We would like to thank James White and Teruki Kamon for useful discussions.
This work has been supported in part by DOE grant DE-FG05-91-ER-40633. The work
of G.P. and X. W. has been supported by a World Laboratory Fellowship.

\newpage
\section*{Figure Captions}
\begin{enumerate}
\item
The correlation between the lightest chargino mass
$m_{\chi^\pm_1}$ and the next-to-lightest neutralino mass $m_{\chi^0_2}$
(top row) for both signs of $\mu$, $m_t=150\GeV$, and (a) the moduli and (b)
dilaton scenarios. Also shown (bottom row) is the absolute value of the
Higgs-mixing parameter $\mu$ versus the gluino mass. Two values of $\tan\beta$
are singled out, larger ones tend to accumulate and are not individually
discernible in the figure.
\label{Figure1}
\item
The first-generation squark and slepton masses as a function of the gluino
mass, for both signs of $\mu$, $m_t=150\GeV$, and (a) the moduli and (b)
dilaton scenarios. The same values apply to the second generation. The
thickness of the lines and their deviation from linearity are because of the
small $\tan\beta$ dependence.
\label{Figure2}
\item
The $\tilde\tau_{1,2}$, $\tilde b_{1,2}$, and $\tilde t_{1,2}$ masses versus
the gluino mass for both signs of $\mu$, $m_t=150\GeV$,
and (a) the moduli and (b) dilaton scenarios. The variability in the
$\tilde\tau_{1,2}$, $\tilde b_{1,2}$, and $\tilde t_{1,2}$ masses is because of
the off-diagonal elements of the corresponding mass matrices.
\label{Figure3}
\item
The one-loop corrected $h$ and $A$ Higgs masses versus the gluino mass for both
signs of $\mu$, $m_t=150\GeV$, and (a) the moduli and (b) dilaton scenarios.
Representative values of $\tan\beta$ are indicated.
\label{Figure4}
\item
The correlated values of $\epsilon_1$ and $\epsilon_b$ (in units of $10^{-3}$)
for both signs of $\mu$, $m_t=130,150,170,180\GeV$, and (a) the moduli and (b)
dilaton scenarios. The ellipses represent the 1-$\sigma$, 90\%CL, and 95\%CL
experimental limits obtained from analyzing all LEP electroweak data.
\label{Figure5}
\item
The parameter space for no-scale $SU(5)\times U(1)$ supergravity (moduli
scenario) in the
$(m_{\chi^\pm_1},\tan\beta)$ plane for (a) $m_t=130\GeV$, (b) $m_t=150\GeV$,
(c) $m_t=170\GeV$, and (d) $m_t=180\GeV$. The periods indicate points that
passed all constraints, the pluses fail the $\bsg$ constraint, the crosses fail
the $(g-2)_\mu$ constraint, the diamonds fail the neutrino telescopes (NT)
constraint, the squares fail the $\epsilon_1-\epsilon_b$ constraint, and the
octagons fail the updated Higgs-boson mass constraint. The reference dashed
line highlights $m_{\chi^\pm_1}=100\GeV$, which is the direct reach of LEPII
for chargino masses. Note that when various symbols overlap a more complex
symbol is obtained.
\label{Figure6}
\item
The parameter space for no-scale $SU(5)\times U(1)$ supergravity (dilaton
scenario) in the
$(m_{\chi^\pm_1},\tan\beta)$ plane for (a) $m_t=130\GeV$, (b) $m_t=150\GeV$,
(c) $m_t=170\GeV$, and (d) $m_t=180\GeV$. The periods indicate points that
passed all constraints, the pluses fail the $\bsg$ constraint, the crosses fail
the $(g-2)_\mu$ constraint, the diamonds fail the neutrino telescopes (NT)
constraint, the squares fail the $\epsilon_1-\epsilon_b$ constraint, and the
octagons fail the updated Higgs-boson mass constraint. The reference dashed
line highlights $m_{\chi^\pm_1}=100\GeV$, which is the direct reach of LEPII
for chargino masses. Note that when various symbols overlap a more complex
symbol is obtained.
\label{Figure7}
\item
The parameter space for the moduli and dilaton $SU(5)\times U(1)$ supergravity
scenarios in the $(m_{\tilde g},\tan\beta)$ plane for $m_t=150\GeV$. The
meaning of the
various symbols is the same as in Figs.~\ref{Figure6},\ref{Figure7}. The
dashed line marks the contour of $m_{\chi^\pm_1}=100\GeV$ ({\em c.f.}
Figs.~\ref{Figure6},\ref{Figure7}) and makes apparent the kinematical
disadvantage of searching for the heavier squarks and gluinos, as opposed to
the lighter charginos.
\label{Figure8}
\item
The parameter space for (a) strict no-scale and (b) special dilaton
$SU(5)\times U(1)$ supergravity in the
$(m_{\chi^\pm_1},\tan\beta)$ plane. The meaning of the
various symbols is the same as in Figs.~\ref{Figure6},\ref{Figure7}.
\label{Figure9}
\item
The number of allowed points in parameter space of the moduli and
dilaton $SU(5)\times U(1)$ supergravity scenarios as a function of $m_t$ when
the basic theoretical and experimental LEP constraints have been imposed
(``theory+LEP"), and when all known direct and indirect experimental
constraints have been additionally imposed (``ALL").
\label{Figure10}
\item
The number of good points in parameter space in the moduli and dilaton
$SU(5)\times U(1)$ supergravity scenarios as a function of $\tan\beta$ for
$m_t=150\GeV$. All known direct and indirect experimental constraints have been
 imposed. Note that for $\mu>0$ there is preference for not so large values
of $\tan\beta$.
\label{Figure11}
\item
The cross section $\sigma(p\bar p\to \chi^\pm_1\chi^0_2X)$ at the Tevatron
(top row) as a function of $m_{\chi^\pm_1}$ for $m_t=150\GeV$ in (a) the moduli
and (b) the dilaton $SU(5)\times U(1)$ supergravity scenarios. Also shown
(bottom row) is the cross section into trileptons, with the 95\%CL
experimental upper limit from CDF as indicated.
\label{Figure12}
\item
The trilepton cross section at the Tevatron  as a function of $m_{\chi^\pm_1}$
for $m_t=150\GeV$ in (a) the strict no-scale and (b) the special dilaton
$SU(5)\times U(1)$ supergravity scenarios. The 95\%CL experimental upper limit
from CDF is indicated.
\label{Figure13}
\item
The still-allowed parameter space in the $(m_{\chi^\pm_1},\tan\beta)$ plane for
$m_t=150\GeV$ in (a) the moduli and (b) the dilaton $SU(5)\times U(1)$
supergravity scenarios. The pluses (+) indicate points explorable with
near-future trilepton searches at the Tevatron, the crosses ($\times$) will
be explorable at LEPII (with ${\cal L}=500\ipb$) through the mixed mode in
chargino pair production, and the diamonds ($\diamond$) will be explorable
at LEPII (with ${\cal L}=500\ipb$) through the dilepton mode in selectron pair
production. Contours of the lightest one-loop corrected Higgs-boson mass are as
indicated (\ie, for $m_h=80,90,100,105,110\GeV$). With
$\sqrt{s}=200\,(210)\GeV$ it should be possible to explore at LEPII up to
$m_h=105\,(115)\GeV$.
\label{Figure14}
\item
The lightest one-loop corrected Higgs-boson mass versus $\tan\beta$ in the
moduli and dilaton $SU(5)\times U(1)$ supergravity scenarios, for
$m_t=150\GeV$. The dotted portions of the vertical lines indicate excluded
ranges of $m_h$. The horizontal line marks the limit of sensitivity of LEPII
with $\sqrt{s}=200\GeV$.
\label{Figure15}
\item
The lightest one-loop corrected Higgs-boson mass versus $\tan\beta$ in the
strict no-scale  (for $m_t=150\GeV$) and the special dilaton (for
$m_t=130,150,155\GeV$) $SU(5)\times U(1)$ supergravity scenarios. The
horizontal line marks the limit of sensitivity of LEPII with
$\sqrt{s}=200\GeV$. Note that LEPII should be able to explore all of the
parameter space in the special dilaton scenario.
\label{Figure16}
\item
The cross section $\sigma(e^+e^-\to Zh)\times f$ versus $m_h$ at LEPII for
$\sqrt{s}=200,210\GeV$ and $m_t=150\GeV$ in the moduli and dilaton $SU(5)\times
U(1)$ supergravity scenarios. Here $f=B(h\to b\bar b)/B(H_{SM}\to b\bar b)$.
Except for the relatively few points deviating from the main curves, the result
is very close to the Standard Model one. The dashed line indicates
the expected level of sensitivity attainable at LEPII.
\label{Figure17}
\item
The branching fraction $B(h\to b\bar b)$ versus $m_h$ for $m_t=150\GeV$ in the
moduli and dilaton $SU(5)\times U(1)$ supergravity scenarios. Note the points
which deviate significantly from the Standard Model expectation (of
$\approx0.85$) owing to the contribution to the total width from the
$h\to\chi^0_1\chi^0_1$ channel.
\label{Figure18}
\item
The Higgs-boson mass contours in the $(m_{\chi^\pm_1},\tan\beta)$ plane for
$m_t=150\GeV$ in the (a) moduli and (b) dilaton $SU(5)\times U(1)$ supergravity
scenarios. The dots represent the still-allowed points in parameter space. For
$\mu<0$ in the dilaton case, the labelling of the mass contours is as for
$\mu>0$.
\label{Figure19new}
\item
The cross section $\sigma(e^+e^-\to\chi^+_1\chi^-_1\to 1l+2j)$ versus
$m_{\chi^\pm_1}$ for chargino searches through the mixed mode at LEPII
($\sqrt{s}=200\GeV$) for $m_t=150\GeV$ in the moduli and dilaton $SU(5)\times
U(1)$ supergravity scenarios. The dashed lines indicate an estimated limit of
sensitivity with ${\cal L}=500\ipb$.
\label{Figure19}
\item
The cross section $\sigma(e^+e^-\to\chi^+_1\chi^-_1\to 2l)$ versus
$m_{\chi^\pm_1}$ for chargino searches through the dilepton mode at LEPII
($\sqrt{s}=200\GeV$) for $m_t=150\GeV$ in the moduli and dilaton $SU(5)\times
U(1)$ supergravity scenarios. The dashed lines indicate an estimated limit of
sensitivity with ${\cal L}=500\ipb$.
\label{Figure20}
\item
The cross sections $\sigma(e^+e^-\to\tilde e\tilde e)$ and
$\sigma(e^+e^-\to\tilde \mu\tilde \mu)$ versus $m_{\chi^\pm_1}$ for selectron
and smuon searches at LEPII ($\sqrt{s}=200\GeV$) for $m_t=150\GeV$ in the
moduli and dilaton $SU(5)\times U(1)$ supergravity scenarios. The dashed lines
indicate an estimated limit of sensitivity with ${\cal L}=500\ipb$. Note that
slepton searches extend the indirect reach of LEPII for chargino masses,
beyond $m_{\chi^\pm_1}\approx100\GeV$.
\label{Figure21}
\item
The total elastic supersymmetric cross section (including selectron-neutralino
and sneutrino-chargino production) at HERA versus $m_{\chi^\pm_1}$ for
$m_t=150\GeV$ in the moduli and dilaton $SU(5)\times U(1)$ supergravity
scenarios. The dashed lines indicate limits of sensitivity with ${\cal L}=100$
and $1000\ipb$.
\label{Figure22}
\item
The value of $B(b\to s\gamma)$ versus the chargino mass for the still-allowed
points in parameter space  with $m_t=150\GeV$ in the moduli and dilaton
$SU(5)\times U(1)$ supergravity scenarios. Wherever possible some values of
$\tan\beta$ have been indicated.
\label{Figure23}
\item
The value of $a^{susy}_\mu$ versus the gluino mass for $m_t=150\GeV$  in the
moduli and dilaton $SU(5)\times U(1)$ supergravity scenarios. The dotted
portions of the curves are excluded. Some values of $\tan\beta$ have been
indicated.
\label{Figure24}

\end{enumerate}


\newpage

\begin{thebibliography}{99}
\bibitem{EriceDec92} For a recent review see \JL, \DVN, and A. Zichichi,
in {\em From Superstrings to Supergravity}, Proceedings of the INFN Eloisatron
Project 26th Workshop, edited by M. J. Duff, S. Ferrara, and R. R. Khuri
(World Scientific, Singapore 1993).
\bibitem{revitalized}I. Antoniadis, J. Ellis, J. Hagelin, and \DVN,
\PLB{194}{87}{231}.
\bibitem{ELN}See \eg, J. Ellis, \JL, and \DVN, \PLB{245}{90}{375};
A. Font, L. Ib\'a\~nez, and F. Quevedo, \NPB{345}{90}{389}.
\bibitem{revamped} I. Antoniadis, J. Ellis, J. Hagelin, and \DVN,
\PLB{231}{89}{65}.
\bibitem{decisive} J. L. Lopez and \DVN, \PLB{251}{90}{73}.
\bibitem{LNY}\JL, \DVN, and K. Yuan, \NPB{399}{93}{654}.
\bibitem{Lacaze}I. Antoniadis, J. Ellis, R. Lacaze, and D.V. Nanopoulos,
Phys. Lett. B {\bf268} (1991) 188.
\bibitem{thresholds} S. Kalara, J.L. Lopez, and D.V. Nanopoulos,
Phys. Lett. B {\bf269} (1991) 84.
\bibitem{sism}S. Kelley, \JL, and \DVN, \PLB{278}{92}{140}; G. Leontaris,
\PLB{281}{92}{54}.
\bibitem{LNZI}\JL, \DVN, and A. Zichichi, \PRD{49}{94}{343}.
\bibitem{IL} L. Ib\'a\~nez and D. L\"ust, \NPB{382}{92}{305}.
\bibitem{KL} V. Kaplunovsky and J. Louis, \PLB{306}{93}{269}.
\bibitem{Ibanez} A. Brignole, L. Ib\'a\~nez, and C. Mu\~noz, FTUAM-26/93
(August 1993).
\bibitem{EN}J. Ellis and \DVN, \PLB{110}{82}{44}.
\bibitem{Lahanas+EKNI+II} J. Ellis, C. Kounnas, and \DVN, \NPB{241}{84}{406},
\NPB{247}{84}{373}; J. Ellis, A. Lahanas, \DVN, and K. Tamvakis,
\PLB{134}{84}{429}.
\bibitem{LN}For a review see A. B. Lahanas and D. V. Nanopoulos,
\PRT{145}{87}{1}.
\bibitem{LNZII}\JL, \DVN, and A. Zichichi, \PLB{319}{93}{451}.
\bibitem{aspects}S. Kelley, \JL, \DVN, H. Pois, and K. Yuan, \NPB{398}{93}{3}.
\bibitem{bsgamma}\JL, \DVN, and G.~T.~Park, \PRD{48}{93}{R974}.
\bibitem{bsg-eps} \JL, \DVN, G.~T.~Park, and A. Zichichi, \PRD{49}{94}{355}.
\bibitem{g-2}\JL, \DVN, and X. Wang, \PRD{49}{94}{366}.
\bibitem{ewcorr}\JL, \DVN, G.~T.~Park, H. Pois, and K. Yuan,
\PRD{48}{93}{3297}.
\bibitem{eps1-epsb} \JL, \DVN, G.~T.~Park, and A. Zichichi, \TAMU{68/93}
(to appear in Phys. Rev. D).
\bibitem{NT} R. Gandhi, \JL, \DVN, K. Yuan, and A. Zichichi, \TAMU{48/93} (to
appear in Phys. Rev. D).
\bibitem{LNWZ}\JL, \DVN, X. Wang, and A. Zichichi, \PRD{48}{93}{2062}.
\bibitem{EF} J. Ellis and G. L. Fogli, \PLB{249}{90}{543}; J. Ellis, G. L.
Fogli, and E. Lisi, \PLB{274}{92}{456}, \PLB{292}{92}{427}, \PLB{318}{93}{154};
F. Halzen, B. Kniehl, and M. L. Stong, Z. Phys. C{\bf58} (1993) 119;
F. del Aguila, M. Martinez, and M. Quiros, \NPB{381}{92}{451}.
\bibitem{oldAmaldi} U. Amaldi, \etal, \PRD{36}{87}{1385}; G. Costa, \etal,
\NPB{297}{88}{244}.
\bibitem{EKN} J. Ellis, S. Kelley and D. V.  Nanopoulos, \PLB{249}{90}{441};
P. Langacker and M.-X. Luo, \PRD{44}{91}{817};
U. Amaldi, W. de Boer, and H. F\"urstenau, \PLB{260}{91}{447};
F. Anselmo, L. Cifarelli, A. Peterman, and A. Zichichi, Nuovo Cim. {\bf104A}
(1991) 1817.
\bibitem{Altarelli} G. Altarelli, in {\em Proceedings of the International
Europhysics Conference on High Energy Physics}, Marseille, France, July 22--28,
1993, edited by J. Carr and M. Perrottet (Editions Frontieres, Gif-sur-Yvette,
1993) CERN-TH.7045/93 (October 1993).
\bibitem{EFL} J. Ellis, G. L. Fogli, and E. Lisi, in preparation.
\bibitem{EWx}L. Ib\'a\~nez and G. Ross, \PLB{110}{82}{215}; K. Inoue, \etal,
Prog. Theor. Phys. 68 (1982) 927; L. Ib\'a\~nez, \NPB{218}{83}{514} and
\PLB{118}{82}{73}; H. P. Nilles, \NPB{217}{83}{366}; J. Ellis, \DVN, and
K. Tamvakis, \PLB{121}{83}{123}; J. Ellis, J. Hagelin, \DVN, and K. Tamvakis,
\PLB{125}{83}{275}; L. Alvarez-Gaum\'e, J. Polchinski, and M. Wise,
\NPB{221}{83}{495}; L. Ib\'a\~nez and C. L\'opez, \PLB{126}{83}{54} and
\NPB{233}{84}{545}; C. Kounnas, A. Lahanas, \DVN, and M. Quir\'os,
\PLB{132}{83}{95} and \NPB{236}{84}{438}.
\bibitem{DL}L. Durand and \JL, \PLB{217}{89}{463}, \PRD{40}{89}{207}.
\bibitem{ANc}P. Nath and R. Arnowitt, \PLB{289}{92}{368}.
\bibitem{LNPWZh}\JL, \DVN, H. Pois, X. Wang, and A. Zichichi,
\PLB{306}{93}{73}.
\bibitem{DNh}M. Drees and M. Nojiri, \PRD{45}{92}{2482}.
\bibitem{LNYdmI}\JL, \DVN, K. Yuan, \NPB{370}{92}{445}.
\bibitem{KLNPYdm}S. Kelley, \JL, \DVN, H. Pois, and K. Yuan,
\PRD{47}{93}{2461}.
\bibitem{KT}See \eg, E. Kolb and M. Turner, {\em The Early Universe}
(Addison-Wesley, 1990).
\bibitem{muproblem}J. E. Kim and H. P. Nilles, \PLB{138}{84}{150} and
\PLB{263}{91}{79}; E. J. Chun, J. E. Kim, and H. P Nilles, \NPB{370}{92}{105}.
\bibitem{Casasmu}J. Casas and C. Mu\~noz, \PLB{306}{93}{288}.
\bibitem{GM}G. Giudice and A. Masiero, \PLB{206}{88}{480}.
\bibitem{Kennedy} D. Kennedy and B. Lynn, \NPB{322}{89}{1};
D. Kennedy, B. Lynn, C. Im, and R. Stuart, \NPB{321}{89}{83}.
\bibitem{PT} M. Peskin and T. Takeuchi, \PRL{65}{90}{964};
W. Marciano and J. Rosner, \PRL{65}{90}{2963};
D. Kennedy and P. Langacker, \PRL{65}{90}{2967}.
\bibitem{efflagr} B. Holdom and J. Terning, \PLB{247}{90}{88};
M. Golden and L. Randall, \NPB{361}{91}{3};
A. Dobado, D. Espriu, and M. Herrero, \PLB{255}{91}{405}.
\bibitem{AB}G. Altarelli and R. Barbieri, \PLB{253}{90}{161}
\bibitem{ABJ}G. Altarelli, R. Barbieri, and S. Jadach, \NPB{369}{92}{3}.
\bibitem{ABC}G. Altarelli, R. Barbieri, and F. Caravaglios, \NPB{405}{93}{3}.
\bibitem{BFC}R. Barbieri, M. Frigeni, and F. Caravaglios, \PLB{279}{92}{169}.
\bibitem{ABCII}G. Altarelli, R. Barbieri, and F. Caravaglios,
\PLB{314}{93}{357}.
\bibitem{Altlecture} G. Altarelli, CERN-TH.6867/93 (April 1993).
\bibitem{D0top} S. Abachi, \etal\ (D0 Collaboration), submitted to Phys. Rev.
Lett. (January 1994).
\bibitem{Thorndike} E. Thorndike, Bull. Am. Phys. Soc. {\bf38}, 922 (1993);
R. Ammar, \etal, CLEO Collaboration, \PRL{71}{93}{674}.
\bibitem{BG}R. Barbieri and G. Giudice, \PLB{309}{93}{86}.
\bibitem{Oshimo} N. Oshimo, \NPB{404}{93}{20}; Y. Okada, \PLB{315}{93}{119}; R.
Garisto and J. N. Ng, \PLB{315}{93}{372}; F. Borzumati, DESY 93-090 (August
1993); M. Diaz, VAND-TH-93-13 (October 1993).
\bibitem{GSW}See \eg, B. Grinstein, R. Springer, and M. Wise,
\NPB{339}{90}{269}; M. Misiak, \PLB{269}{91}{161}, and references therein.
\bibitem{Bertolini}S. Bertolini, F. Borzumati, A. Masiero, and G. Ridolfi,
\NPB{353}{91}{591}.
\bibitem{oldg}J. Bailey \etal, \NPB{150}{79}{1}.
\bibitem{kinoII} For a recent review see, T. Kinoshita, Z. Phys. C{\bf56}
(1992) S80, and in {\em Frontiers of High Energy Spin Physics}, Proceedings
of the 10th International Symposium on High Energy Spin Physics, edited by
T. Hasegawa, N. Horikawa, A. Masaike, and S. Sawada (Universal Academy Press,
1993).
\bibitem{newg}M. May, in AIP Conf. Proc. USA Vol. 176 (AIP, New York, 1988)
p. 1168; B. L. Roberts, Z. Phys. {\bf C56} (1992) S101.
\bibitem{PS85}W.H. Press and D.N. Spergel, Astrophys. J. {\bf 296}
(1985) 679.
\bibitem{Gould}A. Gould,  Astrophys. J. {\bf 321} (1987) 560, 571;
{\bf 328} (1988) 919; {\bf 388} (1992) 338.
\bibitem{KEKII}M. Mori {\it et al.} (Kamiokande Collaboration),
Phys. Lett. B {\bf 289} (1992) 463.
\bibitem{KEK}M. Mori {\it et al.} (Kamiokande Collaboration),
Phys. Lett. B {\bf 270} (1991) 89.
\bibitem{GGR91}G. Gelmini, P. Gondolo, and E. Roulet, Nucl. Phys.
B {\bf 351} (1991) 623.
\bibitem{KAM}M. Kamionkowski, Phys. Rev D {\bf 44} (1991) 3021.
\bibitem{trileptons} J. Ellis, J. Hagelin, \DVN, and M. Srednicki,
\PLB{127}{83}{233}; P. Nath and R. Arnowitt, \MODA{2}{87}{331}; R. Barbieri,
F. Caravaglios, M. Frigeni, and M. Mangano, \NPB{367}{91}{28};
H. Baer and X. Tata, \PRD{47}{93}{2739}; H. Baer, C. Kao, and X. Tata,
\PRD{48}{93}{5175}.
\bibitem{White} Talk given by J. T. White (D0 Collaboration) at the 9th Topical
Workshop on Proton-Antiproton Collider Physics, Tsukuba, Japan, October 1993.
\bibitem{Kato} Talk given by Y. Kato (CDF Collaboration) at the 9th Topical
Workshop on Proton-Antiproton Collider Physics, Tsukuba, Japan, October 1993.
\bibitem{Kamon} T. Kamon, private communication.
\bibitem{Sopczak} A. Sopczak, L3 note 1543 (November 1993).
\bibitem{LG}\JL, \DVN, and X. Wang, \PLB{313}{93}{241}.
\bibitem{HHG}See \eg, {\it The Higgs Hunter's Guide}, J. Gunion, H. Haber, G.
Kane, and S. Dawson (Addisson-Wesley, Redwood City, 1990).
\bibitem{LNPWZ}\JL, \DVN, H. Pois, X. Wang, and A. Zichichi,
\PRD{48}{93}{4062}.
\bibitem{HH}R. Hempfling and A. Hoang, DESY 93-162 (November 1993).
\bibitem{Grivaz}J.-F. Grivaz, LAL preprint 92-64 (November 1992).
\bibitem{Dionisi}C. Dionisi, \etal, in Proceedings of the ECFA Workshop on LEP
200, Aachen, 1986, ed. by A. B\"ohm and W. Hoogland, p. 380.
\bibitem{hera} \JL, \DVN, \XW, and \AZ, \PRD{48}{93}{4029}.
\end{thebibliography}
\end{document}